\documentclass[aps,prb,twocolumn,superscriptaddress,longbibliography]{revtex4-2}
\usepackage[urlcolor=blue, colorlinks=true]{hyperref}
\usepackage{graphicx}
\usepackage{bm}
\usepackage{textcomp}
\usepackage{units}
\usepackage{tabularx}
\usepackage{longtable}
\usepackage{array}
\usepackage{amsmath}
\usepackage{amsfonts}
\usepackage{amssymb}
\usepackage{xspace}
\usepackage{ulem}
\usepackage{booktabs} 
\usepackage{multirow} 
\usepackage{makecell} 
\usepackage{xcolor}
\usepackage{setspace}
\usepackage{placeins}

\newcolumntype{C}{>{\centering\arraybackslash}X}

\newcommand{\pgi}{Peter Grünberg Institut (PGI-3), Forschungszentrum Jülich, 52425 Jülich, Germany}
\newcommand{\jara}{Jülich Aachen Research Alliance (JARA), Fundamentals of Future Information Technology, 52425 Jülich, Germany}
\newcommand{\rwth}{Experimentalphysik IV A, RWTH Aachen University, 52074 Aachen, Germany}
\newcommand{\graz}{Institute of Physics, University of Graz, 8010 Graz, Austria}

\def\degC{$\rm ^{\circ}C$\xspace}
\def\deg{$\rm ^{\circ}$\xspace}

\begin{document}

\title{Tracing the film structure of an organic semiconductor with photoemission orbital tomography}
\author{Monja Stettner} 
    \affiliation{\pgi} \affiliation{\jara} \affiliation{\rwth}
\author{Siegfried Kaidisch}
    \affiliation{\graz}
\author{Andrey V. Matetskiy}
    \affiliation{\pgi} \affiliation{\jara}
\author{Eric Fackelman}
    \affiliation{\pgi} \affiliation{\jara} \affiliation{\rwth}
\author{Serguei Soubatch}
    \affiliation{\pgi} \affiliation{\jara}
\author{Christian Kumpf}
    \affiliation{\pgi} \affiliation{\jara} \affiliation{\rwth} 
\author{Fran\c{c}ois C. Bocquet}
    \affiliation{\pgi} \affiliation{\jara}
\author{Michael G. Ramsey}
    \affiliation{\graz}
\author{Peter Puschnig}
    \affiliation{\graz} 
\author{F. Stefan Tautz} \email{s.tautz@fz-juelich.de}
    \affiliation{\pgi} \affiliation{\jara} \affiliation{\rwth}
   
\date{\today}
\begin{abstract}

Photoemission orbital tomography (POT) is a powerful tool for investigating the orbitals and electronic band structure of oriented layers of organic molecules. In many cases, POT allows conclusions to be drawn regarding the geometric structure, but so far it has been mainly applied to (sub)monolayers and rarely to bilayers, raising the question of whether POT can also provide structure information for thicker films. Here, we use POT to analyze the band dispersion in up to eight layers of $\alpha$-sexithiophene (6T) adsorbed on Cu(110)-p($2\times1$)O. This linear oligomer turns out to be a textbook example that exemplifies the concepts of intra- and intermolecular band dispersion in molecules. Moreover, the rich band and orbital structure information available from POT for this system enables us to trace subtle changes in the crystal structure as a function of layer thickness. Specifically, we find that the periodicity of an intermolecular band changes with film thickness, revealing an increase of the intralayer distance between the molecules with the number of layers. At the same time, the momentum distribution of photoemission from the highest occupied molecular orbital of 6T discloses a decrease of the molecular tilt angle. Following the evolution of tilt angle and lattice constant with layer thickness, we observe---purely based on electronic structure data---that the surface-templated monolayer structure relaxes into the structure of bulk 6T crystals. The experimental findings agree well with the results of density functional theory calculations.

\end{abstract}

\maketitle
\newpage 

\singlespacing
\section{Introduction}

In angle-resolved photoemission spectroscopy (ARPES), the energy and momentum of electrons emitted from a material via the photoelectric effect are analyzed~\cite{Hufner2003}.
Advances in ARPES instrumentation in recent years have enabled efficient measurements of photoelectron intensity distributions over a wide range of binding energies $E_{\mathrm{b}}$ and parallel components $k_x$ and $k_y$ of the momentum~\cite{Sobota2021}.
For organic molecules, photoemission orbital tomography (POT) provides a straightforward interpretation of full ARPES data cubes $I(E_{\mathrm{b}},k_x,k_y)$.
In POT, one relates constant-binding-energy intensity maps measured in the $k_x, k_y$ plane, so-called momentum maps, to calculated Fourier transforms of the real-space molecular orbitals. 
This approach relies on the plane-wave approximation for the photoelectron final state, rendering measured momentum maps proportional to projections of hemispherical cuts through the real-space orbitals' three-dimensional Fourier transform  into the $k_x, k_y$ plane~\cite{Puschnig2009}.
Thereby, momentum maps serve as fingerprints of the orbitals from which the photoelectrons were released, thus enabling the direct observation of the orbital structure~\cite{Puschnig2017}.

In the past, POT was primarily used to study (sub)monolayers of oriented organic molecules \cite{Stadtmueller2012, Wiessner2014,Ules2014, Weiss2015, Hollerer2017, Yang2019PhysChem, Yang2019, Saettele2021,Yang2022,Haags2025}, whereas applications to bi- and multilayers are so far limited to just a few systems~\cite{Stadtmueller2015, Reinisch2016,Grimm2018}.
Although POT holds great promise for investigating how structural and electronic properties of organic layers evolve when transitioning from a monolayer (ML) towards a few layers, it is  challenging to apply POT in this context, because the high orientational order of the molecules in the film must be preserved also for thicker layers.

Using conventional ARPES, $E(k)$ band dispersions of layers of organic molecules have been studied routinely \cite{Hasegawa1994,Rojas2001,Yang2003Intro,Yamane2003,Koller2007,Berkebile2009,Machida2010,Wiessner2013,Vasseur2016}.
Depending on the molecules' geometry and orientation relative to each other, intermolecular band structures as well as intramolecular electronic properties of organic molecules have been observed.
A notable example is the work by Koller \textit{et al.}~\cite{Koller2007}, who demonstrated the possibility to distinguish between intra- and intermolecular dispersion by measuring ARPES  perpendicular and parallel to the long molecular axis of the chain-like molecule sexiphenyl (6P).
Similar findings were observed in thick films of the organic molecules pentacene~\cite{Berkebile2008} and $\alpha$-sexithiophene (6T)~\cite{Berkebile2009}.
For example, a pronounced intermolecular dispersion was observed for a 300\,\AA~thick film of 6T on Cu(110)-p($2\times1$)O~\cite{Berkebile2009}.

However, the electronic structure and the film geometry have not yet been investigated in the interesting range of one to a few layers of 6T, where one expects the evolution of a surface-templated thin-film structure towards the bulk crystal structure.
Here, the full data cube of POT, including both band and momentum maps, offers unique potential: it allows for the determination of lattice periodicities from the intermolecular electronic band structure  \cite{Ules2014,Reinisch2016,Lueftner2017,Bone2023}, and of molecular tilt angles from momentum maps of specific orbitals \cite{Puschnig2009,Reinisch2014,Grimm2018}.\\

In this work, we first present an in-depth conceptual analysis of the electronic band structure of oriented 6T films on Cu(110)-p($2\times1$)O, based on the wealth of information that is available from POT. 
In the second step, we use this insight to derive the in-plane lattice constant and the molecular tilt angle in films of 6T molecules with thicknesses ranging from one to eight ML---notably all from electronic structure data. 
Thereby, we are able to present  a consistent picture of the molecular stack evolving from a surface-templated monolayer to a film with bulk structure. 

\section{Experimental details}
\label{sec:Experimental}
All experiments were performed in an ultra-high vacuum chamber with a base pressure in the 10$^{-10}$\,mbar range.
After repeated cycles of Ar$^+$-sputtering ($p_{\mathrm{Ar}}\,=\,1\times10^{-5}$\,mbar; $E\,=\,1$\,keV; 30\,min) and annealing (600\degC; 30\,min), the Cu(110) substrate was heated to 300\degC (measured with a DIAS Pyrospot DP 10N pyrometer with $\epsilon=$\,0.05) and exposed to oxygen at $p=1\times$10$^{-7}$\,mbar for 10\,min.
This resulted in the well-known p(2$\times$1) oxygen-induced reconstruction, wherein the oxygen atoms are arranged in rows along the [001] direction~\cite{Kishimoto2008}.
This reconstruction emerges as stripes of (2$\times$1)O-reconstructed Cu(110) separated by unreconstructed bare Cu(110).
Increased oxygen exposure yields wider Cu-O stripes~\cite{Cicoira2006}.
However, overexposure to oxygen leads to the formation of another oxygen-induced reconstruction, namely the c(6$\times$2)~\cite{Kishimoto2008}.
Thus, in our experiments, the oxygen exposure was set as high as possible so that no indications of the c(6$\times$2) structure arise in the low-energy electron diffraction (LEED) pattern.

Before use, the 6T molecules (Tokyo Chemical Industry Co. Ltd.) were purified by sublimation. 
The 6T molecules were evaporated from a Kentax three-cell evaporator and adsorbed on the Cu(110)-p($2\times1$)O substrate kept at $-$5\degC.
According to Ref.~\cite{Berkebile20095A}, deposition of a chain-like molecule on a substrate below room temperature leads to a higher nucleation density and, thus, suppresses  the growth of three-dimensional islands.
Note, however, that although the mobility of the molecules is reduced, the substrate's capability to control the molecular orientation remains intact.
The evaporation rate was calibrated with LEED using the findings of Ref.~\cite{BerkebilePhD}.
An evaporation rate of approximately 12\,min per layer was used (T$_{\mathrm{evap.}}$\,=\,250\degC).
After deposition, all experiments were performed at room temperature, increasing molecular diffusion again.

For the POT measurements performed in this work, a HIS14 HD UV source (FOCUS GmbH) with He~I\,$\alpha$  at 21.22\,eV as the main line was used.
Note that the He~I\,$\beta$ satellite emission line at 23.09\,eV contributed 1 to 2 \% to the total radiation.
The focusing mirror and the incident angle (65\deg with respect to the surface normal) led to 30\% s- and 70\% p-polarization of the light, which was considered in the photoemission simulations where necessary.
Photoelectrons emitted in the hemisphere above the sample were analyzed using a NanoESCA MARIS momentum microscope (FOCUS GmbH).
The core components of this instrument are a  photoemission electron microscope (PEEM) and two consecutive hemispherical energy filters.

A three-dimensional data cube $I(E_{\mathrm{b}}$,$k_x$,$k_y$) was obtained by measuring the $k_x$,$k_y$ distribution at different binding energies $E_{\mathrm{b}}\,=\,-\,(E-E_{\mathrm{F}})$, where $E_{\mathrm{F}}$ is the Fermi energy.
Most of the data shown in this paper were recorded with an acquisition time of 10\,s per image, \textit{i.e.}, per energy step in the data cube.
For the tilt angle analysis in section~\ref{sec:Tilt}, however, images were accumulated for 5\,min at fixed binding energies.
The energy resolution of the NanoESCA momentum microscope was set to 220\,meV.

For all measurements, inhomogeneities in the detector gain were corrected by dividing each image by a reference image obtained with secondary electrons, as these are expected to provide an isotropic photoelectron distribution.
Furthermore, the acquired images are not perfectly isochromatic, \textit{i.e.}, the energy within a single image depends slightly on the position $k_x$~\cite{Escher2010,delaPena2010}.
This non-isochromaticity introduces a parabolically-shaped energy distortion in one direction of the field of view~\cite{delaPena2010} and was corrected as described in Ref.~\cite{YangPhD}.
Note that this correction is not possible if only a single momentum map at a fixed binding energy (\textit{i.e.}, not an entire $I(E_{\mathrm{b}}$,$k_x$,$k_y$) data cube) is recorded, as for the analysis shown in Fig.\,\ref{fig:Tilt_analysis}.
However, the parabola used for the correction of the full data cubes can provide an estimate for the energy broadening present in the noncorrected data: 
In the momentum maps of the highest occupied molecular orbital (HOMO), the maximum expected energy difference is 0.07\,eV, which is negligible in the present case, since the width of the HOMO is approximately 0.35\,eV (see, for example, Fig.\,\ref{fig:Intra/Intermolecular_dispersion}).

\begin{figure*}[!htb] 
	\centering
	\includegraphics[width=1\textwidth]{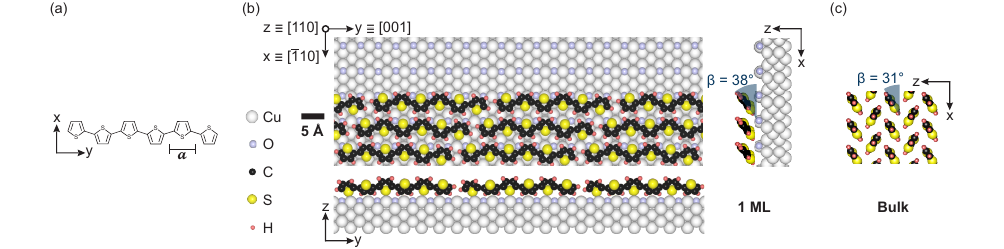}
	\caption{(a) Chemical structure of $\alpha$-sexithiophene (6T). The length $a$ of a single thiophene unit (1T) is indicated. (b) DFT-calculated structure of a monolayer (ML) of 6T adsorbed on Cu(110)-p($2\times1$)O, viewed from the top and along the $x$ ($\equiv [\overline{1}10]$) and $y$ ($\equiv [001]$) directions. In the upper part, the Cu(110)-p($2\times1$)O substrate is shown without molecules. (c) Side view of the 6T bulk structure, as found by Horowitz \textit{et al.}~\cite{Horowitz1995}. The tilt angles $\beta$ of the molecular plane against the surface plane are indicated in panels (b) and (c).}
	\label{fig:Structure}
\end{figure*}

\section{Computational details}
\label{sec:Comp.Details}
We employed density functional theory (DFT) for calculating both film structures and photoemission intensities.  
Four systems were considered: (i) a single 6T molecule without substrate, (ii) a single 6T layer adsorbed on the Cu(110)-p($2\times1$)O substrate, (iii) two ML of 6T on Cu(110)-p($2\times1$)O, and (iv) a free-standing double-layer of 6T molecules in the bulk geometry, in the following referred to as `bulk' 6T.

For the single 6T molecule, system (i), we used the data from the \textsc{Molecular Orbital Database} (Ref.~\cite{Puschnig2020}, entry 16) and computed photoemission momentum maps and band maps using the \textsc{kmap.py} program \cite{kMap}. 
In these calculations, the molecular tilt angle, as experimentally found by Horowitz \textit{et al.}~\cite{Horowitz1995}, was taken into account when simulating the momentum maps.

For systems (ii)-(iv), we performed density functional calculations, employing the Vienna \textit{ab initio} simulation package (VASP)~\cite{kresse1993ab, kresse1996efficiency, kresse1996efficient} and a repeated-slab approach.
For 1\,ML of 6T on Cu(110)-p($2\times1$)O, system (ii), we used the experimental value of $3.595$\,{\AA} for the lattice constant of Cu \cite{PARK2015885}. 
The repeated slab consisted of five atomic layers of Cu including the oxygen-reconstructed uppermost layer.
On top of this, we placed a 6T molecule  in a $1\times7\times1$ supercell. The long molecular axis was aligned parallel to the oxygen rows. The initial tilt angle was set to 39\deg.
Additionally, we added a $20$\,{\AA} vacuum layer, leading to a total slab height of $28.525$\,\AA.
Finally, to avoid spurious interactions between slabs in the $z$ direction, a dipole layer was inserted in all calculations \cite{Neugebauer199216067}.

We used the exchange-correlation functional by Perdew, Burke and Ernzerhof (PBE) \cite{perdew1996generalized}, the Grimme-D3 dispersion correction with the Becke-Johnson damping scheme \cite{grimme2011effect, grimme2010consistent} and the projector augmented wave (PAW) formalism \cite{kresse1994norm, kresse1999ultrasoft}. We employed an
energy cutoff of $400$\,eV and a $\Gamma$-centered $4\times1\times1$ $k$-mesh.
For the ionic relaxations, we applied the conjugate-gradient algorithm and allowed for atomic displacements in all but the lowest two substrate layers.
The converged geometry revealed a molecular tilt of $38^{\circ}$ and is shown in Fig.~\ref{fig:Structure}\,(b).

For optimizing the 2\,ML geometry, system (iii), we started from the relaxed 1\,ML structure [\textit{i.e.}, system (ii)] and added a second-layer molecule to the $1\times7\times1$ supercell. This second molecule was aligned parallel to the first molecule but tilted in the opposite direction compared to the molecules in the first layer, to mimic the herringbone arrangement of the bulk structure of 6T \cite{Horowitz1995}. 
Additionally, we enlarged the height of the unit cell to have again a $20$\,{\AA} vacuum layer separating repeated slabs in the $z$ direction, resulting in a slab height of $35.867$\,{\AA}.
The first step of the geometry optimization aimed at the correct adsorption position of the second layer, especially the molecule's $y$ position, that is, the relative shift along the long molecular axis of the second layer, when compared to the first layer. 
To do this, we froze the substrate and the first molecular layer and treated the second-layer molecule as a rigid body, leaving us with a six-dimensional optimization problem. 
By a series of single-point calculations, the center-of-mass force and torque acting on the rigid molecule were minimized by iteratively translating and rotating the molecule until an energetic minimum was reached.
In a second optimization step, we allowed all internal degrees of freedom of the second-layer molecule to be relaxed and performed a usual VASP geometry optimization, as described in detail above. 
Note, however, that we kept both the substrate and the first molecular layer frozen during this geometry optimization.
This ensured that any resulting changes in the electronic properties of the system can be attributed solely to the presence of the additional layer.
Lastly, we note that, in this geometry optimization, as well as in the above-mentioned single-point calculations, we used the same numerical settings as in the optimization of the 1\,ML system outlined above.

For the relaxed structure of 1\,ML (system (ii)) and 2\,ML (system (iii)) of 6T on Cu(110)-p($2\times1$)O, we simulated photoemission intensity maps using a damped plane wave as photoelectron final state, as described in Ref.~\cite{Lueftner2017}. 
In the underlying VASP calculations, we increased the plane-wave cutoff to 500 eV and used a dense $k$ point sampling of $4 \times 18 \times 4$. 
For the photoemission simulations, we employed a photon energy of 21.2\,eV, a calculated work function of 4.21\,eV for system (ii) (1\,ML) and 4.17\,eV for system (iii) (2\,ML), and, to mimic the finite escape depth of photoelectrons, a damped plane wave with a damping constant of $1$\,\AA$^{-1}$. 
When computing a series of constant-binding-energy momentum maps over a wide binding energy range, the basis of a simulated data cube $I(E_{\mathrm{b}},k_x,k_y)$, we employed artificial broadening of $\Delta E_{\mathrm{b}} = 0.05$\,eV and $\Delta k_x = \Delta k_y =0.1$\,\AA$^{-1}$. 

For system (iv), `bulk' 6T, we computed a free-standing double layer of 6T in the bulk geometry. 
To this end, we started from the crystal structure of 6T as reported in Ref.~\cite{Horowitz1995}, and performed a geometry optimization in a similar manner as described above for systems (ii) and (iii). 
From this optimized structure, we cut out two layers of 6T and added a vacuum slab of 20\,\AA. Without further geometry optimization, we then simulated photoemission intensity maps, using the same approach and settings as described above for systems (ii) and (iii). 
We note that the optimization produced only minor structural differences compared to the structure identified by Horowitz \textit{et al.}\,\cite{Horowitz1995}. 
To render the real-space structure models in this paper, we used the software \textsc{Vesta}~\cite{VESTA}.

\section{Results and discussion}

\subsection{Geometric structure}
$\alpha$-Sexithiophene (6T) is a linear oligomer consisting of six thiophene (1T) rings that are connected by carbon-carbon bonds at their 2- and 5-positions.
Its chemical structure is shown in Fig.\,\ref{fig:Structure}\,(a). In the  gas phase, neighboring thiophene groups are twisted with respect to each other. 
In the twisted configuration, the symmetry group of the molecule is $\bar{\mathrm{I}}$ in the Hermann Mauguin notation (Schoenflies symbol $C_\mathrm{i}$). 
In crystallized form (also in thin films), the molecule tends to planarize, since this allows effective $\pi$-stacking. The symmetry group of planar 6T is $2/{\rm m}$ ($C_\mathrm{2h}$).  

Upon deposition on the Cu(110)-p($2\times1$)O substrate, the 6T molecules uniaxially align with their long molecular axes along the Cu-O rows, \textit{i.e.}, in the $[001]$ direction~\cite{Oehzelt2009}. 
The relaxed structure of a single layer of 6T on Cu(110)-p($2\times1$)O, as calculated by DFT, is shown in Fig.\,\ref{fig:Structure}\,(b).
Throughout the manuscript, we choose the ${[\overline{1}10]}$ direction as the Cartesian $x$ axis and the $[001]$ direction as the $y$ axis. 
Note that, with this convention, the long molecular axis is parallel to the $y$ direction.
As seen in the $xz$ projection in Fig.\,\ref{fig:Structure}\,(b), the molecular planes are tilted with respect to the surface, enabling an overlap of $\pi$ orbitals of neighboring molecules in the $x$ direction.
The bulk structure of 6T, as found by Horowitz \textit{et al.} using x-ray diffraction~\cite{Horowitz1995}, is shown in Fig.\,\ref{fig:Structure}\,(c).
When comparing the tilt angle of the molecules in the first layer (DFT calculation) and in the bulk structure (x-ray diffraction~\cite{Horowitz1995}), a reduction from 38\deg to 31\deg is observed.
This change of the tilt angle is discussed in section~\ref{sec:Tilt}.

\subsection{Intra- and intermolecular dispersion}
\label{sec:Dispersion}

\begin{figure}[!htb]
	\centering
	\includegraphics[width=1\columnwidth]{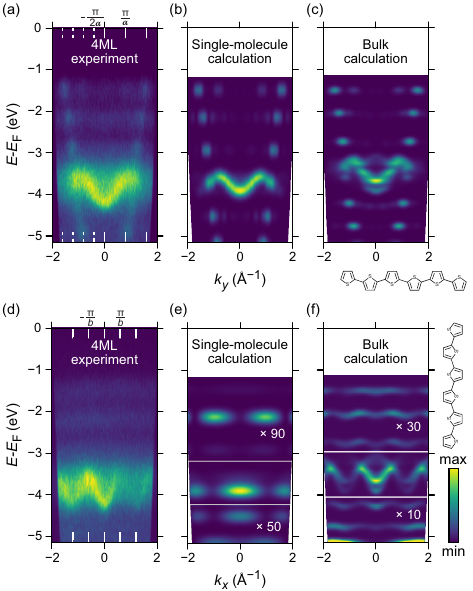}
	\caption{Band maps  (a)-(c) along and (d)-(f) perpendicular to the long molecular axis of 6T for (a),\,(d) the experiment on a 4\,ML 6T film, (b),\,(e) the DFT calculation for a single molecule, and (c),\,(f) the DFT calculation for bulk 6T. $E_{\rm F}$ is the experimental Fermi level. The calculated maps are shifted in energy such that the HOMO is aligned with the experimentally measured HOMO in panel (a). The vacuum level in the single-molecule calculation is at $E-E_{\rm F}=3.3$\,eV.  To aid comparison with the experiment, in the band maps in panels (b), (c), (e) and (f) two rotational domains are considered. In both cases (single molecule and bulk), the tilt angle of the bulk structure ($\beta=\pm 31$\deg against the $xy$ plane) is used. In panels (a) and (d), $k$ vectors corresponding to the lengths $(2)a$ of the (bi)thiophene units (Brillouin zone (BZ) boundaries in case of an infinite polythiophene chain ($\infty$T)) and the intermolecular distance $b$ in the 4\,ML experiment are indicated. The value for $b$ is a result of the analysis in section\,\ref{sec:Coverage} (see Tab.\,\ref{tab:Fitting_results}). In panels (e) and (f), the intensity is increased in parts of the band maps in order to make more molecular levels visible.}
	\label{fig:Intra/Intermolecular_dispersion}
\end{figure}

Fig.~\ref{fig:Intra/Intermolecular_dispersion}\,(a) shows a band map of a 4\,ML 6T film, measured at $k_x=0$ and along the $k_y$ direction in the range $\pm 2$\,\AA$^{-1}$. 
Approximately 4\,eV below the Fermi level, we observe a seemingly continuous band with a bandwidth of $\sim$1\,eV. 
It is cosine-like, with its minimum appearing at $k_y=0$. 
Additionally, we observe a much broader quasiband of two times six nearly equally spaced prolate emissions, symmetrically spaced around $k_y=0$. 
They extend from approximately 1.5 to 5\,eV binding energy. 
Note that the fourth emission from the top is hard to make out, since it overlaps with the continuous band. 
In Fig.~\ref{fig:Intra/Intermolecular_dispersion}\,(d), an $E(k_x)$ band map of the same 4\,ML film, recorded at $k_y=0$ is displayed. 
Again, at binding energy around 4\,eV a continuous cosine-like band, albeit with shorter $k$-space period, is observed. However, there are no signs of the broad quasiband: Instead, dispersionless states at $\sim 2.2$\,eV and, even weaker, at $\sim 1.5$\,eV binding energy are discernible. 
The fact that the band structures in Fig.~\ref{fig:Intra/Intermolecular_dispersion}\,(a) and (d) are symmetric with respect to the $\bar{\Gamma}$ point is a consequence of time-reversal symmetry $k_\parallel \rightarrow -k_\parallel$.    
Note that the asymmetric intensity distribution is caused by the light incidence direction in the experiment.

To interpret the experimental band maps in Fig.~\ref{fig:Intra/Intermolecular_dispersion}\,(a) and (d), we differentiate between intra- and intermolecular dispersion~\cite{Koller2007}.
\textit{Intramolecular dispersion} builds on the fact that, conceptually, the molecular orbitals of 6T can be constructed from the molecular orbitals of a single thiophene (1T) ring. 
When the individual 1T units are connected to form 6T, each molecular orbital of 1T gives rise to six 6T levels that form an intramolecular \textit{quasiband}; its bandwidth is proportional to the hopping integral between the 1T units. 
The discrete levels  within the quasiband arise from combinations of the underlying 1T orbital with an increasing number of nodal planes between them (see below). 
At the same time, in a crystalline molecular film, each of the 6T orbitals is also expected to overlap with the corresponding orbitals of neighboring molecules, additionally giving rise to the \textit{intermolecular dispersion} of conventional bands.

Distinguishing between intra- and intermolecular dispersion in 6T on Cu(110)-p($2\times1$)O becomes possible by measuring band maps either along or perpendicular to the long molecular axis~\cite{Berkebile2009} and comparing them to corresponding calculations.
Since the discrete quasiband in Fig.~\ref{fig:Intra/Intermolecular_dispersion}\,(a) is also predicted by a single-molecule calculation (Fig.\,\ref{fig:Intra/Intermolecular_dispersion}\,(b)), it can only be due to intramolecular interactions.
Moreover, in the bulk calculation in $k_y$ direction (Fig.\,\ref{fig:Intra/Intermolecular_dispersion}\,(c)), both the discrete quasiband and the seemingly continuous band look essentially the same  as in the single-molecule calculation, \textit{i.e.}, there cannot be a significant contribution of intermolecular interaction in the $k_y$ direction. 
In other words, also the seemingly continuous band must be of intramolecular character, which means that we should expect it to consist of discrete lobes spaced so closely that they cannot be separated clearly, both in experiment and theory. 
This conjecture is fully borne out by a more detailed analysis presented in section~\ref{sec:Nonbonding_and_bonding_quasibands}. 

In contrast, in the calculated single-molecule band map perpendicular to the long molecular axis ($k_x$ direction, Fig.\,\ref{fig:Intra/Intermolecular_dispersion}\,(e)), no dispersion is found, while the measurement clearly reveals the existence of a band (Fig.\,\ref{fig:Intra/Intermolecular_dispersion}\,(d)), as mentioned above.  
The bulk calculation (Fig.\,\ref{fig:Intra/Intermolecular_dispersion}\,(f)), however, shows a dispersing band similar to the measurement (Fig.\,\ref{fig:Intra/Intermolecular_dispersion}\,(d)), and therefore, the dispersion in the $k_x$ direction must stem from intermolecular interactions.
We can thus conclude that unlike the intramolecular dispersion in the $k_y$ direction, the band observed in the $k_x$ direction (Figs.\,\ref{fig:Intra/Intermolecular_dispersion}\,(d) and (f)) is a band in the usual sense of the word, \textit{i.e.}, formed from a practically infinite number of molecules and not only from six 1T units, as the quasibands.
Thus, we designate this continuous band along the $k_x$ direction as a conventional intermolecular band in the following.
Thus far, our findings are in agreement with Ref.~\cite{Berkebile2009}, in which a 300\,\AA\,thick film of 6T was investigated using conventional ARPES.

We end this section by briefly discussing three additional features of the calculations that are notable: (i) The intermolecular band in $k_x$ direction is essentially degenerate with the seemingly continuous quasiband in $k_y$ direction (Figs.\,\ref{fig:Intra/Intermolecular_dispersion}\,(c) and (f)), with which it also happens to share a similarly large dispersion.
This degeneracy suggests that both bands originate from the same states.
This conjecture is validated in section~\ref{sec:Intermolecular_band}.  
(ii) In the $k_x$ direction, the bulk calculation also predicts a dispersion for all levels of the broad $k_y$ quasiband (Fig.\,\ref{fig:Intra/Intermolecular_dispersion}\,(f)). 
However, this dispersion is so weak that it cannot be resolved in experiment, where instead we observe faint dispersionless intensity at the corresponding energies (Fig.\,\ref{fig:Intra/Intermolecular_dispersion}\,(d)).  
It is also much smaller than the dispersion of the intermolecular band mentioned in the previous paragraph (Fig.\,\ref{fig:Intra/Intermolecular_dispersion}\,(f)). 
As we will see in section~\ref{sec:Intermolecular_band}, this large difference in $k_x$ dispersion is a consequence of the real-space structure of the orbitals contributing to the respective bands. 
(iii) The $k_x$ positions of the lobes in the  $E(k_x)$ band map for the single-molecule calculation (Fig.\,\ref{fig:Intra/Intermolecular_dispersion}\,(e)) are determined by the presence of two 6T mirror domains on Cu(110)-p($2\times1$)O, in which the molecules have opposite tilt angles of $\pm\,31^\circ$.
This is more clearly visible in the momentum maps in Figs.\,\ref{fig:2Subbands}\,(b) and (d), where the intensity distribution along the stripe-like emissions (\textit{i.e.}, for constant $k_y$) differs for orbitals contributing to the nonbonding and the bonding quasibands. A more detailed discussion is provided in section~\ref{sec:Momentum_signatures_of_the_bonding_and_nonbonding quasibands}.
The strongly reduced intensity of the lobes above and below $-4$\,eV compared to the ones at $-4$\,eV results from the fact that the band maps in Figs.\,\ref{fig:Intra/Intermolecular_dispersion}\,(d)-(f) are cut at $k_y=0$, whereas all orbitals besides the HOMO$-$9 have their maximum intensities at $|k_y| > 0$ (see Fig.\,\ref{fig:2Subbands} below).

\subsection{Nonbonding and bonding quasibands}
\label{sec:Nonbonding_and_bonding_quasibands}

\begin{figure*}[!htb]
	\centering
	\includegraphics[width=1\textwidth]{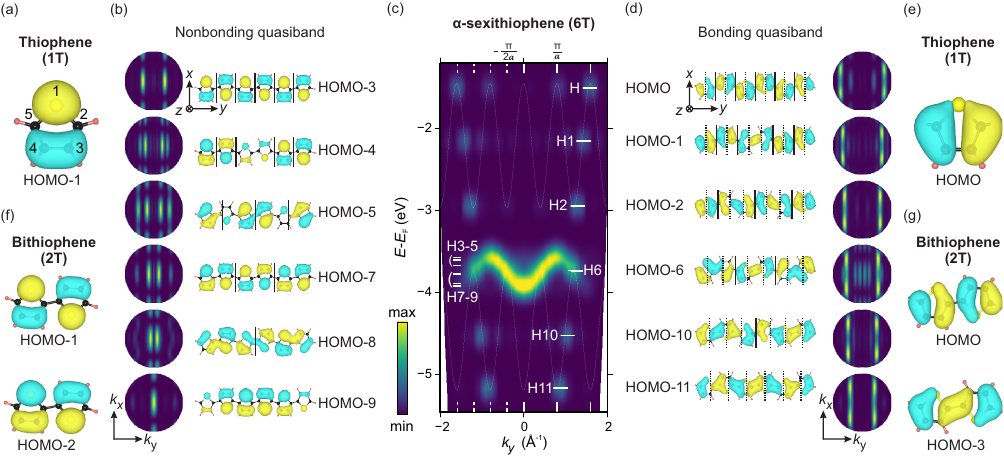}
	\caption{(a) DFT-calculated HOMO$-$1 of a single thiophene unit (1T). The atoms in the ring are numbered to enable clear identification. (b) Momentum maps and wave functions, both derived from a single-molecule DFT calculation,  of the 6T orbitals that contribute to the nonbonding quasiband. These orbitals can be constructed from combinations of a 1T HOMO$-$1 at each unit. Solid vertical lines mark approximate nodal planes, see main text for details. (c) Band map calculated for a single 6T molecule (vacuum level at $E-E_{\rm F}=3.3$\,eV as in Fig.\,\ref{fig:Intra/Intermolecular_dispersion}\,(b)). Orbital energies of HOMO (H) to HOMO-11 (H11) are marked. The sinusoidal lines serve as guides for the eye to highlight the two minibands (see main text for details). Solid (dashed) white marks at $n \pi/a$ ($n\pi/(2a)$) indicate $k_y$ vectors belonging to the 1T (2T) periodicity. (d) Wave functions and momentum maps, both derived from a single-molecule DFT calculation,  of the 6T orbitals that contribute to the bonding quasiband. These orbitals can be constructed from combinations of a 1T HOMO at each unit. Dotted vertical lines mark nodal planes deriving directly from the 1T HOMO, solid vertical lines mark nodal planes arising from the combination of 1T HOMOs at neighboring units. (e) DFT-calculated HOMO of a single thiophene unit (1T). (f) DFT-calculated HOMO$-$1 and HOMO$-$2 of bithiophene (2T). (g) DFT-calculated HOMO and HOMO$-$3 of 2T.  To aid comparison with experiments (in which two rotational domains are always present), the energy and momentum maps in panels (b), (c) and (d) were generated for a pair of single molecules with tilt angles $\beta=\pm 38$\deg against the $xy$ plane.}
	\label{fig:2Subbands}
\end{figure*}

We now look in more detail at the genesis of the two distinct intramolecular quasibands in Figs.\,\ref{fig:Intra/Intermolecular_dispersion}\,(a)-(c).  To this end, we first consider orbitals of both 1T and 6T, as calculated for single isolated  molecules and displayed in Fig.\,\ref{fig:2Subbands}. 
The HOMO$-$1 of 1T shown in Fig.\,\ref{fig:2Subbands}\,(a) has a nodal surface connecting the carbon atoms 2 and 5.
Thus, the 6T orbitals arising from combinations of 1T HOMO$-$1 orbitals generally show a nodal surface along the C$-$C bonds between the thiophene rings, making them nonbonding with respect to chain cohesion.
HOMO$-$3, HOMO$-$4, HOMO$-$5, HOMO$-$7, HOMO$-$8, and HOMO$-$9 of 6T (Fig.\,\ref{fig:2Subbands}\,(b)) belong to this \textit{nonbonding quasiband}. 
Because the overlap (and therefore the hopping integral) between neighboring 1T units in this quasiband is small, it has a relatively small bandwidth ($\sim 0.6$\,eV), making it appear continuous both in the calculated band map in Fig.\,\ref{fig:2Subbands}\,(c) and the experimental one in Fig.\,\ref{fig:Intra/Intermolecular_dispersion}\,(a).

The HOMO of 1T shown in Fig.\,\ref{fig:2Subbands}\,(e), in contrast, has a high electron density at the 2- and 5-positions, albeit of opposite phase. 
If all neighboring 1T HOMO orbitals in a 6-chain of 1T rings are coupled in phase (note the $180^{\circ}$ rotation between adjacent 1T units!), a high electron density is present at the C$-$C bonds between the thiophene rings of 6T, giving rise to the HOMO$-$11 (Fig.\,\ref{fig:2Subbands}\,(d)) at the minimum of another quasiband in 6T. 
In order of ascending energy,  HOMO$-$11, HOMO$-$10, HOMO$-$6, HOMO$-$2, HOMO$-$1 and HOMO are part of this quasiband, which, following the nature of the state at the bottom of this quasiband, we call the \textit{bonding quasiband}. 
Because the lowest orbital in this quasiband (HOMO$-$11) is bonding and the highest one (HOMO) is antibonding between all 1T units, the bandwidth of this quasiband ($\sim 4$\,eV) is substantially larger than that of the nonbonding quasiband. 
The discontinuous nature of the  bonding quasiband (as a quasiband) is therefore conspicuous, and it extends both to lower and higher energies than the nonbonding quasiband. We note that the all-in-phase combination of 1T HOMO orbitals in the bonding quasiband is highest in energy, because head and tail of the 1T HOMO have opposite signs.

Because of time-reversal symmetry, both quasibands are symmetric around $k_y=0$. 
Due to this symmetry, the orbitals in the finite oligomer are standing waves arising from left- and right-travelling waves ($\pm\,k_y$). 
From this point of view, the discreteness of the quasibands can be understood as a consequence of electron confinement in the finite molecular quantum well (see below).  

It is instructive to analyze the nodal structure of the orbitals in Figs.\,\ref{fig:2Subbands}\,(b) and (d) in more detail. 
As mentioned above, the 6T orbitals of the nonbonding quasiband generally show a single nodal surface along the long axis ($y$ axis) of the molecule.
However, it is important to note that the existence of a nodal plane in the 6T states constructed from the 1T HOMO$-$1 is not strictly enforced by symmetry, because the $yz$ plane is not a symmetry plane of 6T.
Indeed, the HOMO$-$5 and the HOMO$-$8 do show some finite wave function amplitude in the $yz$ plane. 
In contrast, the orbitals forming the bonding quasiband do not have $yz$ nodal planes (Fig.\,\ref{fig:2Subbands}\,(d)).
This is an essential difference in comparison to the nonbonding quasiband, which can be traced back to the shapes of the constituent 1T HOMO and 1T HOMO$-$1. 
It has important consequences regarding hybridization between neighboring molecules in $x$ direction, as will be discussed in section~\ref{sec:Intermolecular_band}. 

In Figs.\,\ref{fig:2Subbands}\,(b) and (d), we have also indicated (approximate) nodal planes in the $xz$ plane (thin solid and dotted lines). 
They result from the construction of the 6T orbitals belonging to nonbonding and bonding quasibands from the 1T HOMO$-$1 or 1T HOMO, respectively, and their number increases from zero for the HOMO$-$9 to five for the HOMO$-$3, and from six for the HOMO$-$11 to eleven for the HOMO. 
Again, nodal planes in the $xz$ plane through the center of the molecule are not strictly enforced by symmetry and thus only approximate.
Nevertheless, in the context of constructing the (nonbonding or bonding) intramolecular quasibands of 6T from orbitals of 1T units, we can reasonably focus on the chain-like connectivity in 6T and ignore the $180^\circ$ rotations of neighboring 1T units which breaks the $xz$ mirror symmetry of the molecule. 
The qualitative structure and systematics of the orbitals including their approximate nodal planes in Fig.~\ref{fig:2Subbands} justifies this approach. 
However, for reference we note again that the correct symmetry group of planar 6T is $2/{\rm m}$ ($C_\mathrm{2h}$). 
It therefore has a twofold rotation axis perpendicular to the molecular plane, an inversion center, both in the center of the molecule, and a mirror plane $\sigma_\mathrm{h}$. 
Since all irreducible representations of $2/{\rm m}$ are one-dimensional, all orbitals of 6T have to be either even or odd with respect to these symmetry operations, which one can easily verify in Figs.~\ref{fig:2Subbands}\,(b) and (d). 
In this work, only $\pi$-orbitals, \textit{i.e.}, orbitals that are odd regarding $\sigma_\mathrm{h}$, are considered.

There is a further notable difference between the nonbonding and bonding quasibands. 
The former has its minimum at $k_y=0$, the latter close to $|k_y|=\pi/a$ (which for an infinite polythiophene chain ($\infty$T) would be the Brillouin zone (BZ) boundary), see Figs.~\ref{fig:Intra/Intermolecular_dispersion}(a) and  ~\ref{fig:2Subbands}(c), where $a=3.9$\,\AA\, is the length of a 1T unit (Fig.~\ref{fig:Structure}(a)). The reason is clear: the 1T HOMO orbitals contribute six further nodes to each orbital in the bonding quasiband (dotted lines in Fig.~\ref{fig:2Subbands}(d)) in addition to the nodes of the equivalent orbitals in the nonbonding quasiband (solid lines in Fig.~\ref{fig:2Subbands}(b) and (d)). This can be seen by comparing HOMO$-$11 vs.~HOMO$-$9 etc. We note that the six remaining nodal planes in the 6T HOMO$-$11 (Fig.~\ref{fig:2Subbands}\,(d)) suggest the existence of six additional orbitals at even lower energies, with the number of nodal planes reducing from five to zero (as in the nonbonding quasiband in Fig.~\ref{fig:2Subbands}\,(b)). They can be constructed from the 1T HOMO$-3$ (not shown, see Ref.~\cite{Puschnig2020}, entry 14) in much the same way as discussed above for the bonding and nonbonding quasibands. In fact, since these six orbitals will occur in the range from $k_x\approx 0$ to $k_x\approx \pi/a$, they can also be understood as a part of the bonding quasiband mentioned above. 
We note that for $\infty$T, the then continuous 
band between HOMO$-$11 and HOMO in Fig.\,\ref{fig:2Subbands}(c) could be backfolded around $\pm\pi/a$ into the first BZ (reduced zone scheme). Then, the all-in-phase combination of the 1T HOMO would appear at $k_y=0$, vertically above the all-in-phase combination of the 1T HOMO$-$3, the bottom of the bonding band. Both would share the same Bloch phase factor, but the lattice-periodic parts $u_{k_y}(y)$ of the wave functions (essentially the underlying 1T orbitals) would differ by one node. Since in the finite 6T chain wave functions are not Bloch functions, the orbitals at each 1T unit look slightly different (\textit{i.e.}, no lattice periodicity of $u_{k_y}(y)$), and for this reason HOMO$-$11 to HOMO in Fig.\,\ref{fig:2Subbands}(c) appear at their generic $k_y$ values between $\pm\pi/a$ and $\pm2\pi/a$ and are not backfolded into a reduced zone scheme. 

Using the analogy with a molecular quantum well, it is straightforward to semiquantitatively understand the wave vectors $k_y$ at which 6T orbitals appear in the band maps in Figs.~\ref{fig:Intra/Intermolecular_dispersion}(a) and  ~\ref{fig:2Subbands}(c). For a particle in the box, the wavelength of a state with $n$ nodes ($n=0,1,2,\dots$) is $\lambda=2L/(n+1)$, where $L$ is length of the box (molecule). The corresponding wave vectors are $k_y=\pm2\pi/\lambda=(n+1)\pi/(Ma)$, where $M=6$ (Fig.~\ref{fig:Structure}(a)). Accordingly, in the nonbonding quasiband the state with $n=5$ (HOMO$-$3) should appear at $k_y=\pm\pi/a$, which it does, while the state with $n=0$ (HOMO$-$9) is expected at $k_y=\pm\pi/(6a)$. However, looking at Fig.~\ref{fig:2Subbands}(c), we observe the bottom of the nonbonding quasiband at $k_y=0$, because the strongest Fourier component of the HOMO$-$9, which has nodes only at the end points of the molecule, must be located at $k_y=0$.
For the bonding quasiband, the state at the top of the band ($n=11$, HOMO) is expected and found at $k_y=\pm2\pi/a$, whereas the state with $n=6$ (HOMO$-$11) is anticipated at $k_y=\pm 1.17\pi/a$. Indeed, the emission of the HOMO$-$11 appears in both experiment (Fig.\,\ref{fig:Intra/Intermolecular_dispersion}\,(a)) and theory (Fig.\,\ref{fig:2Subbands}\,(c)) beyond $|\pi/a|$, but at a slightly lower value $|k_y|\approx 1.07\pi/a$. 
A similar effect was observed for pentacene\,\cite{Berkebile2008} and can be attributed to a termination effect: At both ends of the oligomer the wave function spills out of the molecular quantum well, leading to a smaller overall period in reciprocal space.
Note that the nonbonding quasiband is (almost) unaffected by this termination effect, because of the (approximate) $yz$ nodal plane in its orbitals restricting the spillout at the ends of the molecule.

Finally, we turn to the relative positions of the nonbonding and bonding quasibands. In both experiment and calculation, the average energies of the bonding and nonbonding quasibands are shifted with respect to each other, as indicated by the white solid lines in Figs.\,\ref{fig:MomMaps}\,(a) and (b). In the calculation, this shift is approximately 0.4\,eV (Fig.\,\ref{fig:MomMaps}\,(a)), which is fully explained by the energy difference between constituent 1T orbitals of both quasibands, the HOMO and HOMO$-$1, respectively \cite{Puschnig2020}. However, the experimentally measured band map in Fig.\,\ref{fig:MomMaps}\,(b) shows that the average energy of the bonding quasiband is $\sim0.7$\,eV above that of the nonbonding quasiband. Taking the center of the nonbonding quasiband as a reference, the dashed white line in Fig.\,\ref{fig:MomMaps}\,(b) indicates where the center of the bonding quasiband would be expected if only the energy difference between the underlying 1T orbitals was considered. Thus, we observe an additional upward shift of the bonding relative to the nonbonding quasiband, by approximately 0.3\,eV. We assign this to an effect described by R.\,Hoffmann~\cite{Hoffmann1988}: Due to overlap between constituent orbitals at neighboring sites, antibonding states are generally more antibonding than the bonding ones are bonding, leading to an asymmetric spread of a band around its nonbonding reference level. In Fig.\,\ref{fig:MomMaps}\,(b), this effect is not only apparent from the mentioned additional upward shift of the whole bonding quasiband, but also from the larger differences between the antibonding levels of the bonding quasiband (red, magenta, cyan) compared to the differences between the bonding levels (green, orange, gray, see section~\ref{sec:Momentum_signatures_of_the_bonding_and_nonbonding quasibands} for the placement of these color bars). We note that while a similar asymmetric spread in a bonding quasiband has been noticed for 6P in Ref.\,\cite{Koller2007} (where the orbitals forming the bonding and nonbonding quasibands, HOMO and HOMO$-$1 of a single benzene ring, are degenerate), it is, remarkably, not observed in the calculation (see Fig.\,\ref{fig:MomMaps}\,(a)).

\subsection{Bonding and antibonding minibands}
\label{sec:Bonding_and_antibonding_minibands}

In both experiment (Fig.~\ref{fig:Intra/Intermolecular_dispersion}(a)) and theory (Fig.~\ref{fig:2Subbands}(c)) we observe weak replicas of the 6T HOMO at $k_y=\pm\pi/a$. 
Theory even shows similar replicas for other orbitals of the bonding quasiband, with decreasing intensity from HOMO to HOMO$-$11.
The reason is clear: Because of the alternating position of sulfur atoms in 6T (see Fig.~\ref{fig:Structure}\,(a)), the correct monomer of 6T is---strictly speaking---bithiophene instead of thiophene.
The doubling of the repeat distance from $a$ to $2a$ reduces the wave vector associated with each orbital by half. Indeed, the up/down alternation of the lobe pairs in the 6T HOMO clearly reveals this. 
In the case of $\infty$T, the doubling of the periodicity from 1T to 2T monomers would lead to backfolding the band structure into the smaller BZ of the 2T monomer, by reflection of the bands at $\pm\pi/2a$. 
Such backfolding creates band crossing points at the new zone boudaries $\pm\pi/2a, \pm 3\pi/2a, ...$. If the bands were continuous, gaps would be opening at the crossing points, \textit{i.e.}, around $E-E_{\rm F}\approx -3.5$\,eV.  
In Fig.~\ref{fig:2Subbands}(c), we have therefore plotted two \textit{minibands} (thin white lines) with a gap between them, instead of the original bonding quasiband. The state above the gap is the HOMO$-$2, while the HOMO$-$6 lies directly below the gap.

Since the minibands are associated with the 2T periodicity, it should also be possible to construct the corresponding 6T orbitals from suitable orbitals of 2T, in the same way as this was done above for the 1T monomer. 
Indeed, the orbitals in the upper miniband can be constructed exclusively from the 2T HOMO (Fig.\,\ref{fig:2Subbands}\,(g)): the 6T HOMO is the all-in-phase combination, while in the 6T HOMO$-$2 signs alternate. 
Since the 2T HOMO is antibonding between the two thiophene units, we refer to the upper miniband as the \textit{antibonding miniband}. In this miniband, again the all-in-phase combination is highest in energy, because head and tail of the 2T HOMO have opposite signs, as is the case for the 1T HOMO (see section~\ref{sec:Nonbonding_and_bonding_quasibands} above). 
We note in passing that the state in the center of the antibonding miniband, the 6T HOMO$-$1, which needs alternating in-phase and out-of-phase combinations of the 2T HOMO, requires for its construction a partitioning of 6T into 2T units that leaves half a 2T at either end; only by this shifting of the 2T lattice by half a lattice constant, the required number of in-phase and out-of-phase contacts is reconcilable with the requirement (following from the particle-in-the-box picture) to have for this orbital an antinode in the center of the molecule. Clearly, the way in which 6T is partitioned into 2T units is arbitrary, as the extension to an $\infty$T chain from which a 6T oligomer is cut out reveals.     

Similarly, orbitals in the lower miniband follow from the 2T HOMO$-$3  (Fig.\,\ref{fig:2Subbands}\,(g)). The 6T HOMO$-$11 is the all-in-phase combination, while in the 6T HOMO$-$6 signs alternate.
Since the 2T HOMO$-$3 is bonding between the thiophene units, the lower miniband is referred to as the \textit{bonding miniband}. 

For an $\infty$T chain with continuous bands, the states at the bottom of the upper miniband and at the top of the lower miniband are standing waves, shifted by half a 2T lattice constant against each other, and the different positions of their nodes and antinodes relative to the 2T lattice are the origin of the energy gap at the BZ boundary. 
In fact, in the finite 6-chain, the 6T HOMO$-$6 and 6T HOMO$-$2 are precursors of these two phase-shifted standing waves. To see this, we consider the bonding and antibonding minibands of $n$T in the limit $n\rightarrow \infty$. As the $n$T chains get longer, the discrete orbitals move closer to each other. 
This will also decrease the gap at  $k_y=\pm\pi/(2a)$ (note that for $n\rightarrow \infty$ the difference in the number of nodes between the states at the upper and lower gap edges becomes irrelevant).
Yet, the nature of the $n$T orbitals at the bottom of the upper (antibonding) miniband and at the top of the lower (bonding) miniband remain distinct: The former stays a purely bonding combination of only the 2T HOMO, while the latter continues to be a purely antibonding combination of exclusively the 2T HOMO$-$3. 
This abrupt, discontinuous change from exclusively 2T-HOMO$-$3-derived to exclusively 2T-HOMO-derived $n$T orbitals in effect leads to a shift of nodes by half a 2T lattice constant (black vertical lines in Fig.~\ref{fig:2Subbands}(c)) and thus to a finite gap even in the $\infty$T system, in which the bands otherwise become continuous. 
Note that the states in these continuous bands of the infinite chain generically are travelling waves (no Bragg scattering) with wavelengths that are not matched to the 2T periodicity. 
But evidently, in finite 6T scattering at the chain ends leads to the formation of standing waves via superposition of states with $\pm k_y$, even within the bands. 
We remark in passing that Fig.~\ref{fig:2Subbands} (c) reveals a regular, non-inverted gap between the two minibands, in which local orbitals with an even (`$s$-like') motif (2T HOMO$-$3) are located energetically below those with an odd (`$p$-like') motif (2T HOMO). 
As a consequence, no Shockley-like local (`surface') state is expected at the ends of long $n$T chains. 
This would require an inverted gap.

We end this section by a final observation regarding the construction of 6T orbitals from either 1T or 2T orbitals. 
Although all orbitals in Fig.\,\ref{fig:2Subbands}\,(d)  can in principle be constructed from 2T component orbitals as well as from 1T ones, a glance at the resultant 6T orbitals reveals why in the band maps in Fig.\,\ref{fig:2Subbands}\,(c) and Fig.\,\ref{fig:Intra/Intermolecular_dispersion}(a) the feature at $k_y=\pm2\pi/a$ is stronger than the one at $k_y=\pm\pi/a$, and why for the other 6T orbitals in the bonding quasiband the backfolded features are hardly visible in the experiment: Their overriding repeat pattern is the one of the 1T monomer, while the 2T monomer appears only as a small perturbation, most evidently in the 6T HOMO. 
In fact, for the nonbonding quasiband, the 1T monomer is even more dominant.
This is a consequence of the shape of the 2T orbitals from which the nonbonding quasiband can be constructed, \textit{i.e.}, the 2T HOMO$-$1 (upper miniband) and HOMO$-$2 (lower miniband), see Fig.\,\ref{fig:2Subbands}\,(f). 
Essentially, the lobes of these orbitals are copies of the 1T HOMO$-$1, arranged in straight lines that do not follow the zig-zag of the 2T units in 6T, thus firmly establishing 1T as the relevant monomer.

\subsection{Momentum signatures of the bonding quasiband} 
\label{sec:Momentum_signatures_of_the_bonding_and_nonbonding quasibands}

In this section, we discuss the momentum maps of the orbitals that belong to the bonding quasiband (bonding and antibonding minibands). For reference, we first look at the momentum maps of the single-molecule calculation in Fig.\,\ref{fig:2Subbands}\,(b) and (d). 
They are dominated by the stripe-like lobes along the $k_x$ direction, which are symmetrically placed around the $k_y=0$ line and disperse within the quasibands to increasing $|k_y|$.
As mentioned briefly in section~\,\ref{sec:Dispersion} in the context with Fig.\,\ref{fig:Intra/Intermolecular_dispersion}\,(e), there is a notable difference in the momentum maps of the orbitals that contribute to either the nonbonding or bonding quasibands. 
For the states forming the nonbonding quasiband, superposition of the two tilted molecular orientations ($\pm\,31^\circ$) shifts the major emission, which is centered at positions $|k_x| > 0$ for untilted molecules, to spread around $k_x = 0$ (see Fig.\,\ref{fig:2Subbands}\,(b)). Conversely, the major lobes of the states forming the bonding quasiband, for untilted molecules centered at $k_x = 0$, are moved to finite values $|k_x| > 0$ (see Fig.\,\ref{fig:2Subbands}\,(d)). 
For reference, the momentum maps of the untilted molecules are available in entry 16 of Ref.\,\cite{Puschnig2020}.
For several orbitals, we also observe relatively strong minor lobes. Notable examples are HOMO$-$6, HOMO$-$5, and HOMO$-$4. 

\begin{figure}[p]
	\centering
    \includegraphics[width=0.915\columnwidth]{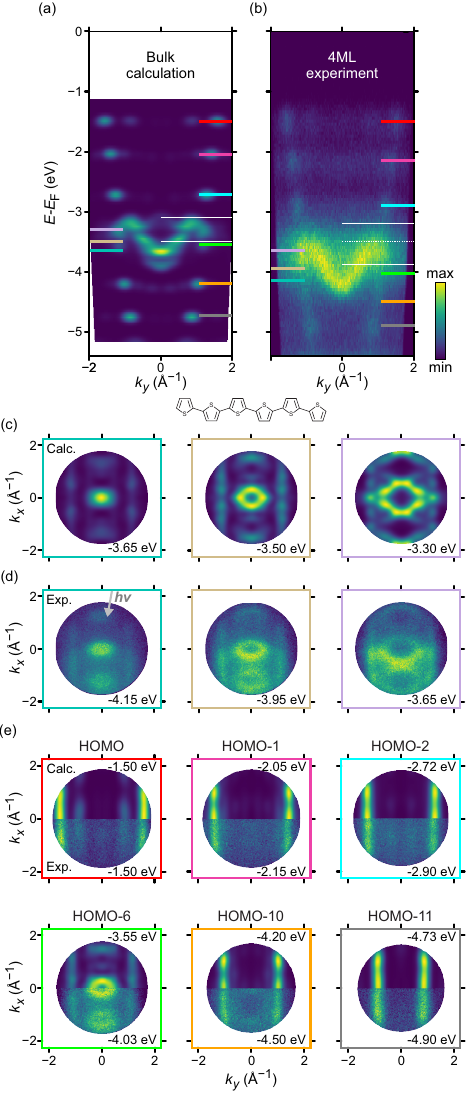}
	\caption{(a) DFT-calculated (for bulk 6T) and (b) measured (for the 4\,ML film) band maps. The calculated band map was energy-shifted such that the HOMOs in experiment and theory align. White solid lines mark the centers of the quasibands. The white dashed line marks the position where the center of the bonding quasiband (relative to the nonbonding quasiband) would be expected if only the energy difference between the underlying 1T orbitals (HOMO and HOMO$-$1, single-molecule calculation) was considered.  (c) Calculated (bulk) and (d) measured (4\,ML) momentum maps, recorded at the energies marked by colored lines in the left margins of panels (a) and (b). In the leftmost map in panel (d), a gray arrow indicates the projection of the light direction into the $k_x$,$k_y$ plane. (e) Direct comparison of calculated (bulk, top) and measured (4\,ML, bottom) momentum maps taken at the energies marked by the colored lines in the right margins of panels (a) and (b).}
	\label{fig:MomMaps}
\end{figure}

The typical stripe-like lobes are also observed experimentally for the orbitals of the bonding quasiband.
This is shown in Fig.\,\ref{fig:MomMaps}\,(e), where the measurement results are compared to momentum maps of the bulk calculation.
To enable a meaningful comparison of momentum maps between experiment and theory, we must align the band structures on the energy axis. DFT calculations typically do not yield accurate ionization energies\,\cite{Puschnig2015}. However, in systems in which the plane-wave approximation for the final state is good, DFT calculations are very reliable with regard to the photoemission intensity distribution in momentum space, provided momentum patterns at suitably chosen energies are compared. In  Fig.\,\ref{fig:MomMaps}\,(a) the calculated band map for the bulk is therefore shifted in energy to align the theoretical HOMO energy with the experimental one in Fig.\,\ref{fig:MomMaps}\,(b). 
For the antibonding miniband (HOMO, HOMO$-$1, HOMO$-$2), the energies at which the cuts through the experimental data cube need to be made to reveal the appropriate momentum map are clear (see Fig.\,\ref{fig:MomMaps}\,(b)).  However, for the bonding miniband (HOMO$-$6, HOMO$-$10 and HOMO$-$11), the experimental lobes in Fig.\,\ref{fig:MomMaps}\,(b) are blurred into each other, such that an accurate determination of the proper experimental binding energy becomes difficult. Therefore, we used the maps themselves to identify the energies at which the expected features in the map are strongest \cite{Puschnig2011,Haags2025}. The respective energies are marked at the right margin of Fig.\,\ref{fig:MomMaps}\,(b). In Fig.\,\ref{fig:MomMaps}\,(e), the calculation for the bulk structure (top) is shown in direct comparison to the experimental maps (bottom). This representation easily allows one to observe the excellent agreement between experiment and theory.  For all orbitals except the HOMO$-$6, only the typical vertical lines appear, \textit{i.e.}, the momentum maps look essentially like the single-molecule calculation in Fig.\,\ref{fig:2Subbands}\,(d). Thus, we can conclude that there is no intermolecular dispersion for this band, in agreement with theory (Fig.\,\ref{fig:Intra/Intermolecular_dispersion}\,(f)). In contrast, the momentum maps of the HOMO$-$6 and the ones taken in the energy range of the nonbonding quasiband (Figs.\,\ref{fig:MomMaps}\,(c) and (d)) strongly differ from the single-molecule calculation (Figs.\,\ref{fig:2Subbands}\,(b) and (d)). As these orbitals are degenerate with the intermolecular band in $k_x$ direction, we may suspect the formation of an $E(k_x,k_y)$ band. This is discussed in section~\ref{sec:Intermolecular_band}.

\subsection{Intermolecular band}
\label{sec:Intermolecular_band}

We now take a closer look at the intermolecular band in $k_x$ direction. In particular, we  focus on its relation to the nonbonding quasiband in $k_y$ direction, with which it is degenerate (see Figs.\,\ref{fig:Intra/Intermolecular_dispersion}\,(a) and (d)). To this end, it is convenient to analyze  experimental momentum maps and compare them to calculated ones. At the left margins of Figs.\,\ref{fig:MomMaps}\,(a) and (b), three energies are marked, at equivalent position with respect to the respective band dispersion in theory and experiment. The corresponding experimentally measured momentum maps are displayed in Fig.\,\ref{fig:MomMaps} (d) alongside the maps from the bulk calculation in Fig.\,\ref{fig:MomMaps} (c). 
We indeed observe excellent agreement between theory and experiment, taking into account that in the experiment the direction of the light incidence, indicated by the gray arrow in the leftmost map in Fig.\,\ref{fig:MomMaps}\,(d), leads to suppression of photoemission in the backward direction, while in the calculated maps the polarization factor has been left out for clarity---these maps therefore have up/down symmetry. In the leftmost maps, a sharp spot at the $\bar\Gamma$ point is visible.
Also, weak replicas of this spot in the next BZs  at $|k_{x}|\approx \pm 1.2$\,\AA$^{-1}$ are discernible, more clearly in the experiment. Moving higher in energy, each spot  opens into an ellipse; in the rightmost panel, the ellipses even reach the BZ boundary in $k_x$ direction and touch each other. The continuity of the ellipses shows that the intramolecular $k_y$ dispersion of the nonbonding quasiband and the intermolecular $k_x$ dispersion are cuts through the same object---a continuous band $E(k_x,k_y)$ over the two-dimensional $k_x,k_y$ plane.  

\begin{figure}[!htb]
	\centering
	\includegraphics[width=0.93\columnwidth]{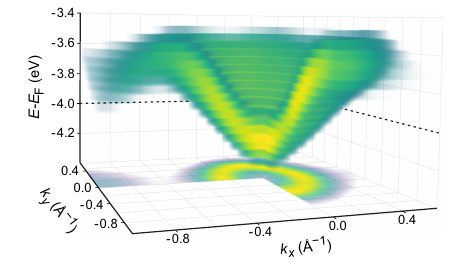}
	\caption{3D representation of the $E(k_x,k_y)$ band measured for a 4\,ML film. To generate this representation, the data of the full $I(E_\mathrm{b}, k_x,k_y)$ data cube in the relevant energy range were processed as follows: First, each $k_x$,$k_y$ map was symmetrized. Then, pre-smoothing was performed in 3D by applying a Gaussian filter. Next, a curvature filter was applied using an analogous formula to Eq.\,(14) in Ref.\,\cite{Zhang2011}, but in three dimensions (see Appendix~\ref{Appendix:3D_curvature}). The necessary derivatives were obtained using Savitzky-Golay filters~\cite{SavGol1964,2020SciPy}. Finally, all intensity values below 35\% of the maximum intensity were removed in order to reduce the background and only focus on the $E(k_x,k_y)$ band. The following parameters were used: Pre-smoothing: $\sigma_{k_x} = \sigma_{k_y} = 10$ and $\sigma_E = 1$; Curvature filtering: $C_k = 0.05$\,\AA$^{-2}$, $C_E = 0.01$\,eV$^2$, window length and polynomial order in the Savitzky-Golay filter~\cite{SavGol1964,2020SciPy}: 100 and 2 for the $k$ axes, and  20 and 2 for the $E$ axis. The opacity as well as the color depends on the intensity values. The opacity of a single data point varies from 0 to 0.5 and the color from dark blue to yellow. To enable a view into the band, data points with negative $k_x$ and $k_y$ values are not shown. A cut through the dataset at $E\,-\,E_{\mathrm{F}}\,=\,-4.0$\,eV (dotted lines) is shown in the $k_x$,\,$k_y$ plane at $E\,-\,E_{\mathrm{F}}\,=\,-4.4$\,eV. Note that, in contrast to the $E(k_x,k_y)$ band, the opacity for the $k_x$,\,$k_y$ map ranges from 0 to 1. The layered structure in vertical direction is a consequence of the finite energy steps in the measured data cube. While mainly the first BZ is shown in the figure, at $k_y$\,=\,0 and $k_x\,<\,-0.6$\,\AA$^{-1}$ the edge of the $E(k_x,k_y)$ band in the neighboring BZ is visible as well.}
	\label{fig:3D_Plot}
\end{figure}

A 3D representation of this band, based on the full experimental photoemission data cube $I(E_\mathrm{b}, k_x, k_y)$,  is shown in Fig.\,\ref{fig:3D_Plot}. Its dispersion in $k_y$ ($k_x$) direction is due to intramolecular (intermolecular) coupling. Notably, the electron delocalization in $x$ direction fundamentally changes the appearance of the momentum maps: whereas the maps of the contributing orbitals (depicted in Fig.\,\ref{fig:2Subbands}\,(b)) exhibit long  stripes in $k_x$ direction, we now observe an ellipse in the $k_x$,$k_y$ plane instead (see Fig.\,\ref{fig:MomMaps}\,(d) and Fig.\,\ref{fig:3D_Plot}). The contraction of the features in $k_x$ direction clearly indicates a delocalization in $x$ direction. We note that the lines along $k_x$ at $|k_{y}|\approx\pm$1\,\AA$^{-1}$ in Fig.\,\ref{fig:MomMaps} (c) and (d) do not belong to the $E(k_x,k_y)$ band---they originate from the HOMO$-$6, which is degenerate with it, see Figs.\,\ref{fig:2Subbands}\,(c) and (d). 

Judging from the maximum slopes of the band dispersions in $k_x$ and $k_y$ directions in Fig.\,\ref{fig:3D_Plot} (cf.~also Figs.\,\ref{fig:Intra/Intermolecular_dispersion}\,(a) and (d)), the intermolecular dispersion is $\sim 1.5$ times steeper than the intramolecular dispersion in the nonbonding quasiband. Remarkably, electrons thus move with a larger phase velocity $ v\,=\,\hbar^{-1} \partial E/\partial k$ between the molecules than within. This is consistent with the elliptical shape of the $E(k_x,k_y)$ band in the right and middle columns of Fig.\,\ref{fig:MomMaps}\,(d). Yet, the overall bandwidth of the intra- and intermolecular dispersion is essentially the same---after all, both contribute to the same band. The larger period in $x$ direction ($b\approx 5$\,\AA, see section \ref{sec:Coverage}) than in $y$ direction ($a=3.9$\,\AA), together with the steeper dispersion in $k_x$, give rise to almost identical bandwidths in $k_x$ and $k_y$ directions.

The $E(k_x,k_y)$ band in Fig.\,\ref{fig:3D_Plot} points to the limits of the orbital picture for oligothiophenes on Cu(110)-p($2\times1$)O in the binding energy range in question. Delocalization within and between the molecules is comparable, such that it seems more appropriate to think of a 2D lattice of 1T units. For a side-by-side arrangement of $\infty$T chains, this would give rise to an energetically sharp single band with the overall shape of the one shown in Fig.\,\ref{fig:3D_Plot}. For the finite oligomer 6T, the confinement of the electrons in the $y$ direction breaks this single band into six transverse subbands, one for each of the contributing orbitals in Fig.\,\ref{fig:2Subbands}(b); these six subbands give rise to the nonbonding quasiband in $k_y$ direction. Unfortunately, we do not resolve the different subbands in Fig.\,\ref{fig:3D_Plot}, but observe a single broadened band instead.

A quantitative analysis of the bandwidth of the $E(k_x,k_y)$ band yields  $\sim0.6$\,eV (see Table~\ref{tab:Fitting_results} below). The geometric arrangement of chain-like molecules (significantly tilted and with their long molecular axis parallel to each other) enables a large $\pi$-orbital overlap between neighboring molecules, leading to this large bandwidth. A similar behavior was observed for a submonolayer of heptacene (0.5\,eV for HOMO$-$1) \cite{Bone2023},  a 200\,\AA~thick film of 6P (0.7\,eV for the intermolecular band) \cite{Koller2007} and a 300\,\AA~thick layer of 6T (0.5\,eV for the intermolecular band) \cite{Berkebile2009}. Note that literature mostly reports smaller values for the band dispersion of single orbitals of organic molecules: For instance, 0.35\,eV for the HOMO of BTQBT \cite{Hasegawa1994}, 0.2\,eV for the HOMO of PTCDA \cite{Yamane2003} and 0.4\,eV for the HOMO of rubrene \cite{Machida2010}. 

\begin{figure}[!t]
	\centering
	\includegraphics[width=0.93\columnwidth]{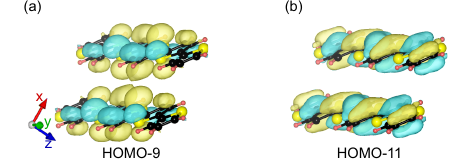}
	\caption{Isosurfaces of (a) HOMO$-$9 (nonbonding) and (b) HOMO$-$11 (bonding) orbitals in the bulk geometry. In both panels the same level of the isosurfaces is shown.}
	\label{fig:Com_Overlap}
\end{figure}

As discussed in the previous section, the intermolecular dispersion of the bonding quasiband is negligable.  This is a consequence of the shape of the corresponding orbitals in real space. 
The orbitals of the bonding quasiband, having the maxima of their probability density in the central $yz$ plane (Fig.\,\ref{fig:2Subbands}\,(d)), cannot sustain an effective overlap between neighboring tilted molecules, in contrast to the orbitals of the nonbonding quasiband (Fig.\,\ref{fig:2Subbands}\,(b)), which have the maxima of their probability density closer to the periphery of the molecules.
This is shown exemplarily for the HOMO$-$9 (nonbonding quasiband) and the HOMO$-$11 (bonding quasiband) in Fig.~\ref{fig:Com_Overlap}.
For the orbitals of the nonbonding quasiband, the intermolecular orbital overlap allows sufficient hybridization between neighboring molecules for a dispersive band to form. 
On the basis of this argument we may also conjecture that the LUMO of 6T, having two nodal planes parallel to the long axis of the molecule (see entry 16 in Ref.\,\cite{Puschnig2020}), will also be able to hybridize between neighboring molecules, leading to a dispersive conduction band in 6T films. 
This is expected to have important consequences for the lowest-energy exciton in 6T films on Cu(110)-p($2\times1$)O, for which the hole, being located in the HOMO, would be much more confined to a single molecule than the electron in the LUMO, which could spread out over several molecules. 
In fact, this is precisely what is observed~\cite{ExcitonPaper}. 

\subsection{Intermolecular dispersion and film structure}
\label{sec:Coverage}
\begin{figure*}[!htb]
	\centering
	\includegraphics[width=1\textwidth]{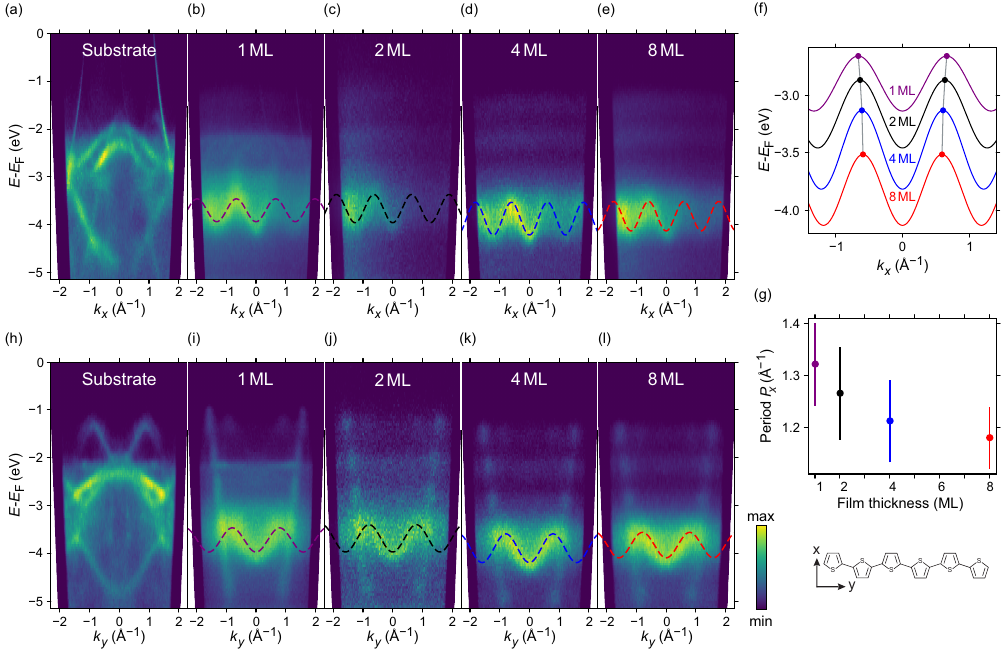}
	\caption{(a)-(e) Band maps measured along the $k_x$ direction ($[\overline{1}10]$ direction of the substrate) for (a) the bare Cu(110)-p($2\times1$)O substrate and (b)-(e) film thicknesses of 1\,ML to 8\,ML of 6T on Cu(110)-p($2\times1$)O as indicated. The dashed lines are cosine fits of the intermolecular dispersion. (f) Comparison of the fitted cosine bands of the intermolecular dispersion from panels (b)-(e) for different film thicknesses, with maxima marked to highlight the change in the period $P$. For better readability, the curves of 1\,ML (purple), 2\,ML (black) and 4\,ML (blue) are vertically shifted by 0.8\,eV, 0.5\,eV and 0.4\,eV, respectively. (g) Period $P_x=2\pi/b$ of the intermolecular dispersion as a function of film thickness. (h)-(l) Band maps measured along the $k_y$ direction ($[001]$ direction of the substrate), for (h) the bare Cu(110)-p($2\times1$)O substrate and (i)-(l) film thicknesses of 1\,ML to 8\,ML of 6T on Cu(110)-p($2\times1$)O as indicated. The dashed lines are cosine fits of the nonbonding quasiband.}
	\label{fig:Fits}
\end{figure*}
In the 6T bulk, molecules form a layered structure, with their long axes aligned along the $y$ direction in the layer plane ($xy$ plane). 
The molecules are also tilted by the  same angle.
From layer to layer (in $z$ direction), the tilt angle alternates between $+31$\deg and $-31$\deg. 
This structure is shown in Fig.\,\ref{fig:Structure}\,(c)~\cite{Horowitz1995}. 

On the Cu(110)-p($2\times1$)O surface, the 6T molecules align uniaxially along the oxygen rows~\cite{Oehzelt2009}.
Our DFT calculations yield a tilt angle of 38\deg for the first layer (Fig.\,\ref{fig:Structure}\,(b)), suggesting that its structure is slightly modified from that of bulk 6T by the presence of the substrate. 
Oehzelt \textit{et al.} studied the geometry of 6T on Cu(110)-p($2\times1$)O, using the near-edge x-ray absorption fine structure (NEXAFS) technique. 
They found the tilt angle to be reduced from 45\deg for 1\,ML to 41\deg for a multilayer~\cite{Oehzelt2009}.
While the absolute values of the tilt angle found with NEXAFS differ from the experimental result for the bulk structure (31\deg) \cite{Horowitz1995} and our DFT prediction for the first layer (38\deg), the tendency (smaller tilt angle for larger film thickness) is the same.

Given the structure of 6T films, it is plausible that the variation of the tilt angle with film thickness is accompanied by a change of their intermolecular distance in $x$ direction. 
This would also explain the mechanism by which the presence of the substrate increases the tilt angle in the 1\,ML film: By the templating effect of the CuO rows, the surface forces the molecules closer to each other, and the molecules respond to this external constraint by tilting further out of the surface plane.

Since POT in principle can quantify both the intermolecular distance (by its impact on the intermolecular dispersion---discussed in the following) and the molecular tilt angle (by its influence on the momentum maps, which in the plane-wave approximation are projections of hemispherical cuts through the oriented 3D Fourier transform of the real-space orbital into the sample surface plane---discussed in section~\ref{sec:Tilt}), it is very well suited to study quantitatively the (coupled) evolution of these two structural parameters as a function of film thickness. 

Moreover, if the photon energy is chosen judiciously (such that photoelectron kinetic energies are close to the minimum of the universal curve of electron inelastic mean free path vs.~kinetic energy), even single-layer resolution can essentially be achieved. 
Since our experiments with He~I light yield photoelectron kinetic energies in the range of 12\,-\,17\,eV, the inelastic mean free path is of the order of 10\,\AA, corresponding to roughly two 6T layers. 
We note that the main contribution to the electrons reaching the analyzer, however, originates from the topmost layer. 
Hence, we expect only a small contribution of the substrate for 1\,ML films, while no substrate features should be observable in the band maps measured at larger film thickness, and contributions of layers below the top layer should be sufficiently damped to enable the measurement of film thickness-dependent data with essentially single-layer resolution.  

In Figs.~\ref{fig:Fits}\,(b)-(e) and (i)-(l), data measured  on samples with four different film thicknesses (1\,ML, 2\,ML, 4\,ML, and 8\,ML) are presented. The data of the 4\,ML film were already presented in the discussion of intra- and intermolecular dispersion in sections~\ref{sec:Dispersion} to \ref{sec:Intermolecular_band}.
For reference, results for the bare Cu(110)-p($2\times1$)O substrate are displayed in Figs.~\ref{fig:Fits}\,(a) and (h). 
The onset of the Cu $d$ band is  observed as a sharp edge at $\sim -2$\,eV. 
While this edge is still clearly visible for the 1\,ML 6T film, Figs.~\ref{fig:Fits}\,(b) and (i), all distinct band features of the Cu(110)-p($2\times1$)O surface are already significantly damped out. 
Moreover, in Fig.~\ref{fig:Fits}\,(i) the lobes of the bonding quasiband are blurred by the substrate. From 2\,ML thickness onwards, the overall appearance of the band maps does not change any more, and in particular, no influence of the substrate is discernible.

\setlength{\tabcolsep}{0.8em}
{\renewcommand{\arraystretch}{1.7} 
\begin{table*}[!htb]
  \caption{\label{tab:Fitting_results} Results of the dispersion fits in Fig.\,\ref{fig:Fits} and the orientation fits in Fig.\,\ref{fig:Tilt_analysis}. $W_y$: bandwidth of the dispersion in $k_y$ direction; $E_{{\rm min},y}$: band minimum in $k_y$ direction; $W_x$: bandwidth of the dispersion in $k_x$ direction, $E_{{\rm min},x}$: band minimum in $k_x$ direction; $P_x$: band period in $k_x$ direction; $b$: intermolecular distance in $x$ direction; $\beta$: molecular tilt angle against the $xy$ plane (absolute value). The fitted full width at half maximum of the Gaussian broadenings on the energy axis range from 0.5 to 0.7\,eV.}
  \begin{tabularx}{\textwidth}{lccccccc}
    \toprule \midrule
    \multirow{3}{*}{ } & \multicolumn{2}{c}{\makecell{$k_y$ direction \\ (nonbonding quasiband)}} & \multicolumn{4}{c}{\makecell{$k_x$ direction \\ (intermolecular band)}} & \multicolumn{1}{c}{\makecell{ ($k_x,k_y$) plane \\ (HOMO)}} \\
    \cmidrule(lr){2-3} \cmidrule(lr){4-7} \cmidrule(lr){8-8}
     & \makecell{ $W_y$\,(eV)} &
    \makecell{$E_{{\rm min},y}$\,(eV)} &
    \makecell{ $W_x$\,(eV)} &
    \makecell{$E_{{\rm min},x}$\,(eV)} &
    \makecell{ $P_x$ \,(\AA$^{-1}$)} &
    \makecell{ $b$\,(\AA)} &
    \makecell{ $\beta$\,(\deg)} \\
    \midrule
    1\,ML & $0.50\pm0.08$  & $-3.97\pm0.08$ & $0.48\pm0.08$ & $-3.94\pm0.08$ & $1.32\pm0.08$ & $4.8 \pm 0.3$ & \\
    2\,ML & $0.6\pm0.1$  & $-3.98\pm0.09$ & $0.6\pm0.2$  & $-4.0\pm0.1$ & $1.27\pm0.09$ & $4.9 \pm 0.4$ & 37\,$^{+\,3}_{-\,2}$ \\
    4\,ML & $0.60\pm0.08$  & $-4.19\pm0.09$ & $0.7\pm0.1$  & $-4.2\pm0.1$ & $1.21\pm0.08$  & $5.2 \pm 0.4$ & 32\,$\pm$\,2\\
    8\,ML & $0.54\pm0.08$  & $-4.10\pm0.08$  & $0.62\pm0.06$  & $-4.13\pm0.07$ & $1.18\pm0.06$ & $5.3 \pm 0.3$ & 31\,$\pm$\,2\\
    \midrule \bottomrule
  \end{tabularx}
\end{table*}

First, we focus on the intermolecular $k_x$ dispersion in Figs.~\ref{fig:Fits}\,(b)-(e). 
Its period $P_x = 2\pi/b$ is related to the intermolecular distance $b$. To see whether the latter changes with film thickness, we fitted the bands with a cosine function on the $k_x$ axis and a Gaussian broadening (in the order of $\thicksim$\,0.6\,eV) on the energy axis. 
The resulting cosine functions are shown as dashed lines in Figs.~\ref{fig:Fits}\,(b)-(e) and in a direct comparison for the different film thicknesses in Fig.~\ref{fig:Fits}\,(f). 
The complete set of fitting parameters for the intermolecular $k_x$ dispersion is summarized in the central part of Table~\ref{tab:Fitting_results}.

The period of the intermolecular dispersion decreases with increasing film thickness, as is clearly visible in Fig.~\ref{fig:Fits}\,(g). 
A reduction from $P_x=(1.32\,\pm\,0.08)$\,\AA$^{-1}$ to $(1.18\,\pm\,0.06)$\,\AA$^{-1}$ corresponds to an increase of the intermolecular distance from $b=(4.8\,\pm\,0.3)$\,\AA~to $(5.3\,\pm\,0.3)$\,\AA.  
Even though the absolute values are smaller than expected, the trend is in agreement with the expectation: For the first ML, the intermolecular distance is governed by the separation between neighboring oxygen rows (5.11\,\AA~in the Cu(110)-p($2\times1$)O  reconstruction), while in the bulk structure the intermolecular distance is 5.52\,\AA~\cite{Horowitz1995}. 
Since we can exclude a calibration error of the $k$ axes, we attribute the smaller absolute values obtained in the fit to the presence of the HOMO$-$6 in the energy range of the fitted dispersion.  
This systematically distorts the results for the period towards higher values.  
Nevertheless, we can conclude that in our experiment the 6T structure is compressed in the $x$ direction by the presence of the substrate and relaxes toward the bulk structure with increasing film thickness.

\begin{figure*}[!htb]
	\centering
	\includegraphics[width=1\textwidth]{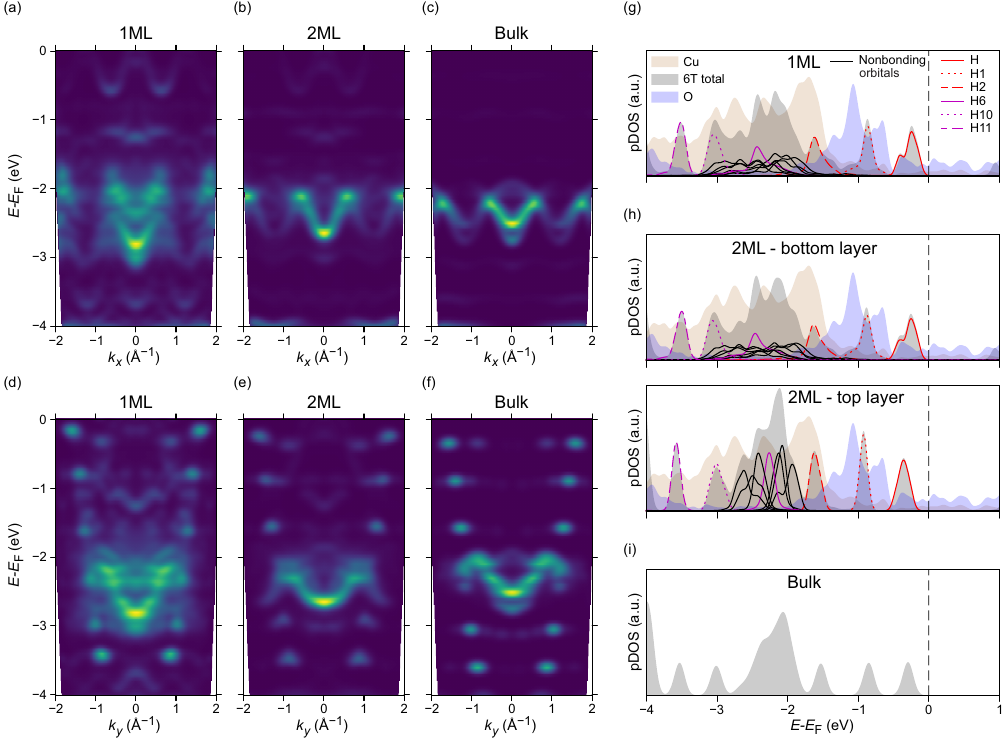}
	\caption{(a)-(c) DFT-calculated band maps in $k_x$ direction. The 1\,ML and the 2\,ML calculations were done on the oxygen-reconstructed Cu(110)-p(2x1)O substrate with an exponentially damped plane-wave photoelectron final state, while in the bulk calculation a double-layer of 6T in the bulk geometry and a plane-wave final state were employed, see section~\ref{sec:Comp.Details} for details. (d)-(f) Same as panels (a)-(c), but in $k_y$ direction. (g)-(i) Projected densities of states (pDOS) for 1\,ML, 2\,ML and bulk 6T. The total contribution of 6T to the density of states is shown by the gray-shaded area, while lines show the respective contributions of each orbital as indicated in the figure legend. Orange and gray shading mark Cu and oxygen contributions, respectively. The pDOS for the 2\,ML calculation (h) is shown separately for the molecules of the bottom layer (\textit{i.e.}, in contact with the substrate) and the top-layer molecules. We note that the overall too small binding energies in the simulations compared to experiment are a result of the DFT treatment with a semi-local exchange-correlation functional. In Fig.\,\ref{fig:Appendix_1ML_DFT} in Appendix~\ref{Appendix_B}, the 1\,ML data of panels (a), (d), and (g) are displayed on a common energy axis, thus aiding the visual assignment of bands to pDOS features.}
	\label{fig:DFT_comparison}
\end{figure*}

The significance of the above result is confirmed by analyzing the $k_y$ dispersion in the same way. Here, we indeed obtain the same period $P_y=1.9$\,\AA$^{-1}$ for all film thicknesses. This constant value should reflect the interring spacing of 6T, \textit{i.e.}, the distance $a=3.9$\,\AA~  
between neighboring thiophene rings. 
However, this chemically defined distance, which cannot be changed significantly through the adsorption, yields a period of only $P_y=2\pi/a=1.59$\,\AA$^{-1}$. 
Thus, similar to the case of the $k_x$ dispersion, it appears as if the presence of the nondispersing HOMO$-$6 distorts the fit towards a larger period in $k$ space. For further analysis, we therefore fixed the period of the $k_y$ dispersion to 1.59\,\AA$^{-1}$.
Then, the only remaining fit parameters are the bandwidth $W_y$ and energy $E_{{\rm min},y}$ (see Table~\ref{tab:Fitting_results}), as well as the Gaussian broadening on the energy axis. 
The resulting cosine functions, shown as dashed lines in Figs.~\ref{fig:Fits}\,(i)-(l), fit the data very well.  

Turning to the fitted bandwidths in Table~\ref{tab:Fitting_results} , one might intuitively expect the bandwidth $W_x$ of the $k_x$ dispersion, being of intermolecular origin, to decrease with increasing intermolecular distance (\textit{i.e.}, larger film thickness). 
Instead, we observe an initial increase (up to 4\,ML) and then a decrease, although the error intervals for the different film thicknesses overlap substantially. 
On the one hand, this result may indicate that besides the intermolecular distance, there are other relevant parameters that influence the bandwidth and change with film thickness, \textit{e.g.}, the tilt angle of the molecules determines the $\pi-\pi$ overlap between neighboring molecules and might, depending on the actual shape of the orbitals, counteract the band narrowing with increasing distance. 
On the other hand, a more careful analysis of the shape of the band provides a possible explanation for the fact that the fitted $W_x$ initially seems to rise before falling again. 
In this context, it is revealing that in Fig.\,\ref{fig:Fits}\,(b) for 1\,ML we observe additional intensity above the maxima of the cosine fit ($\sim$ -\,3~eV; $\sim$ $\pm\,0.7$\AA$^{-1}$)---apparently, the latter does not capture the  full width of the band.

To understand the origin of this additional feature, we look at the calculated $k_x$ dispersion in Fig.\,\ref{fig:DFT_comparison}\,(a). 
For the 1\,ML film, the band gives the impression of being split. 
Its lower section exhibits a clear cosine shape, while the upper part is dominated by intense features above the maxima of the lower segment. 
As a consequence, fitting the experiment with a cosine function will necessarily yield a smaller amplitude than the full bandwidth.
At the same time, in the calculation, the overall bandwidth drops monotonously from 1\,ML to the bulk (Figs.\,\ref{fig:DFT_comparison}\,(a)-(c)). 
According to Figs.\,\ref{fig:DFT_comparison}\,(g)-(i), this is a consequence of two unrelated mechanisms. 
First, there is a clear narrowing of the cosine band from the 2\,ML film to the bulk. 
This is visible both by comparing  the band maps in Figs.\,\ref{fig:DFT_comparison}\,(b) and (c) and juxtaposing the molecule-projected partial density of states (pDOS) of 6T (gray shaded areas) for the 2\,ML top layer (Fig.\,\ref{fig:DFT_comparison}\,(h), bottom) and the bulk (Fig.\,\ref{fig:DFT_comparison}\,(i)). 
Thus, in the calculation, the increased distance between the molecules does lead to a smaller overlap, effects of the changing tilt angles notwithstanding. 
Notably, this tendency is also visible in the experiment, where the fitted $W_x$ for 8\,ML is smaller than for 4\,ML (Table~\ref{tab:Fitting_results}).
Secondly, in addition to the narrowing of the band for thicker layers caused by dwindling intermolecular interactions, there is also a massive influence of the substrate on the first layer.
Apparently, for molecules in direct contact with the substrate, the orbitals of the nonbonding quasiband are not only prone to hybridize with their neighbors, but also with the substrate. 
As before, this can be rationalized by their shapes, having lobes close to the periphery of the molecule. 
This hybridization can be seen as the strong broadening of the pDOS (black lines) for the 1\,ML film in Fig.\,\ref{fig:DFT_comparison}\,(g) and for the bottom layer of the 2\,ML film in Fig.\,\ref{fig:DFT_comparison}\,(h) (we note in passing that the pDOS of the orbitals of the bonding quasiband (red and purple lines) are much sharper, thus hybridizing much less with the substrate).  
We suggest that the fact that this hybridization, although leading to a strong broadening of the band, cannot be captured by the cosine fit, causes the initial rise of the fitted $W_x$.
As the contact layer is covered with the second and further layers, the bandwidth first rises (compare the lower section of the 1\,ML band in Fig.\,\ref{fig:DFT_comparison}\,(a) with the 2\,ML band in Fig.\,\ref{fig:DFT_comparison}\,(b)), before the fits can pick up the narrowing caused by the change of the film structure (compare the 2\,ML band in Fig.\,\ref{fig:DFT_comparison}\,(b) with the bulk band in Fig.\,\ref{fig:DFT_comparison}\,(c)). 
Our result for the bandwidth $W_x$ agrees with the findings of Berkebile \textit{et al.}~\cite{Berkebile2009}, who reported a bandwidth in the order of 0.5\,eV for a 300\,\AA\,film.

Next, we briefly discuss the bandwidth $W_y$ of the $k_y$ dispersion. 
As expected from the intramolecular nature of the dispersion in this direction, there is essentially no dependence of the bandwidth on the film thickness, as a comparison between Figs.\,\ref{fig:DFT_comparison}\,(e) and (f) shows. 
In Fig.\,\ref{fig:DFT_comparison}\,(d) we also observe a strong broadening of the nonbonding quasiband caused by hybridization with the substrate; apparently, interaction with the substrate increases the hopping between 1T units, partially reducing the nonbonding character of the constituent orbitals.
The experimental band map in Fig.\,\ref{fig:Fits}\,(i) also reveals the problem of capturing the full width of the band by the cosine fit. As for the $k_x$ direction, this can explain a small initial rise of $W_y$ (Table~\ref{tab:Fitting_results}). 
In agreement with theory, beyond 1\,ML there is no strong change in the fitted bandwidths $W_y$. 

We finally note that $E_{{\rm min},x}$ and $E_{{\rm min},y}$ agree within the error bars for each coverage, showing that the dispersions in $k_x$ and $k_y$ direction belong to the same $E(k_x,k_y)$ band, as already discussed in section~\ref{sec:Intermolecular_band}.
Regarding the band minima, however, the agreement between experiment and theory is only moderate. 
In experiment, we do not observe a clear tendency in the energy position of the band minimum relative to the HOMO energy (see Figs.\,\ref{fig:Fits}\,(i)-(l)), while in theory the nonbonding quasiband moves closer to the HOMO (Figs.\,\ref{fig:DFT_comparison}\,(d)-(f)) with increasing film thickness. 

\subsection{Orbital momentum maps and film structure }
\label{sec:Tilt}

While dispersive bands permit to derive the distance between molecules, the simultaneous presence of a band without dispersion, for which the momentum maps are essentially the ones of the single molecule, allows the determination of the molecular orientation \cite{Puschnig2009,Reinisch2014,Grimm2018}. 
Therefore, we now turn to the bonding quasiband. In particular, we look at the HOMO at the top of the antibonding miniband. 

Fig.\,\ref{fig:Tilt_analysis}\,(a) shows momentum distribution curves (MDCs) for different film thicknesses (solid lines), extracted from experimental momentum maps of the HOMO that are displayed in the lower halves of the panels in Fig.\,\ref{fig:Tilt_analysis}\,(c). 
Thin black lines in the experimental momentum maps indicate the region of 0.3\,\AA$^{-1}$ width from which the MDCs in Fig.\,\ref{fig:Tilt_analysis}\,(a) were extracted.
As a function of film thickness, the MDCs clearly reveal a shift of the maximum intensity toward smaller $|k_x|$, particularly evident in Fig.~\ref{fig:Tilt_analysis}\,(b).

\begin{figure}[!htb]
	\centering
	\includegraphics[width=1\columnwidth]{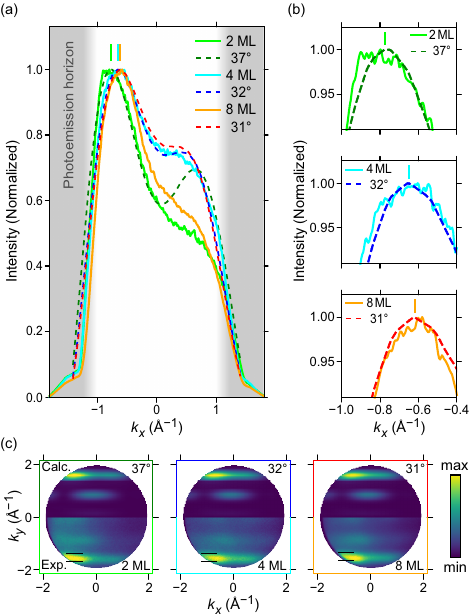}
	\caption{(a) Momentum distribution curves (MDCs) extracted from simulated (dashed lines) and measured (solid lines) momentum maps (both displayed in panel (c)) of the 6T HOMO. (b) Enlarged views around the maxima of the MDCs. To determine the precise positions of the maxima, first a Savitzky-Golay filter~\cite{SavGol1964,2020SciPy} was applied, then the first derivative was calculated, and finally the zero crossing point was found by fitting a straight line to the first derivative. The such-determined zero crossing points were taken as the maxima and, for the experimental MDCs, marked by colored lines in panels (a) and (b).  (c) Momentum maps calculated (top) for  tilt angles 37\deg, 32\deg and 31\deg (single-molecule DFT calculation) and measured (bottom) on the 2\,ML, 4\,ML, and 8\,ML samples as indicated. The momentum maps were measured at $E-E_{\mathrm{F}}= -1.35$\,eV, $-1.55$\,eV, and $-1.45$\,eV for  2\,ML, 4\,ML and 8\,ML, respectively. The simulated MDCs were extracted at $k_{y}=1.56$\,\AA$^{-1}$, while from the measured momentum maps, the data in a range of $\pm\,0.15$\,\AA$^{-1}$ around $k_{y}=-1.51$\,\AA$^{-1}$, $k_{y}=-1.52$\,\AA$^{-1}$, and $k_{y}=-1.50$\,\AA$^{-1}$ were used for 2\,ML, 4\,ML, and 8\,ML, respectively. These intervals are marked by two thin black lines in the respective experimental momentum maps.}
	\label{fig:Tilt_analysis}
\end{figure}

To rationalize this behavior, we simulated HOMO momentum maps for a single molecule with tilt angles ranging from 29\deg to 44\deg in 1\deg steps.
For direct comparison with experiment, these calculated HOMO maps include the polarization factor for the light incidence geometry that was used in the experiment. 
From each theoretical map, we extracted the MDC at $k_{y}=1.56$\,\AA$^{-1}$. 
Using the position of the maximum as the criterion, we then matched the experimental momentum maps for different film thicknesses with calculated maps for different tilt angles. 
MDCs for the best matching angles are  plotted as dashed lines of the corresponding color in Figs.\,\ref{fig:Tilt_analysis}\,(a) and (b), while the corresponding theoretical maps are displayed in the upper halves of Fig.\,\ref{fig:Tilt_analysis}\,(c)  in direct comparison to the assigned experimental maps. 

We find tilt angles of (37\,$^{+\,3}_{-\,2}$)\deg, (32\,$\pm$\,2)\deg and (31\,$\pm$\,2)\deg for 2\,ML, 4\,ML and 8\,ML, respectively (see rightmost column in Table~\ref{tab:Fitting_results}). 
These values are in excellent agreement with the optimized geometry from the DFT calculation for 1\,ML (38\deg) and with the bulk structure determined by Horowitz \textit{et al.}~\cite{Horowitz1995} using x-ray diffraction (31\deg).
The experimental error quoted for the 2\,ML sample is asymmetric because the peak is slightly cut at the photoemission horizon of the experiment. 
Thus, the maximum might occur at a larger tilt angle.
We also note that the analysis of the molecular tilt angle is only carried out for film thicknesses of 2\,ML or higher. 
We found that the tilt angle cannot be unambiguously determined from the HOMO momentum map of the 1\,ML measurement, for two reasons: First, the orbitals contributing to the bonding quasiband are not clearly separated in energy (see Fig.\,\ref{fig:Fits}\,(i)), and second, the 1\,ML measurement is too severely influenced by substrate features (see Figs.\,\ref{fig:Fits}\,(b) and (i)).

\section{Conclusion and outlook}

We have presented an investigation of templated thin films of the linear thiophene oligomer $\alpha$-sexithiophene on the oxygen-reconstructed copper substrate Cu(110)-p($2\times1$)O, using both photoemission orbital tomography and density functional theory. 
Crucially, we have been able to deduce key structural parameters of these films from their measured electronic structure, specifically band dispersion and momentum maps.  

Regarding the electronic structure, we observed a rich band structure, comprising both intermolecular and intramolecular aspects. 
The latter was analyzed as a textbook example of how bands arise when identical units are coupled with varying strengths. 
Specifically, we observed two discrete quasibands, one with strong coupling (giving rise to a broad bonding quasiband), the other with weak coupling between the units (yielding a narrow nonbonding quasiband), in full accord with Berkebile \textit{et al.}~\cite{Berkebile2009}. 
Three remarkable observations in this context are: First, the bands and orbitals can best be understood with a single thiophene as an elementary unit, although neighboring thiophenes are rotated by 180$^\circ$ and thus the correct monomer of $\alpha$-sexithiophene is bithiophene.
As it turns out, the backfolding induced by the correct periodicity produces only very faint additional intensity in the band structure, and hence a single quasiband affords the better description than two split minibands. 
This demonstrates that when constructing the bands, it is not so much the geometric structure of the molecular backbone which determines the most expedient elementary unit, but rather the shape of the specific unit's orbital in question. 
In this respect, one needs to differentiate between bands that originate from different orbitals (in the present case, the bonding quasiband exhibits a stronger effect of backfolding compared to the nonbonding quasiband). 
This notwithstanding, it is important to remember that photoemission experiments modulate the $k$-resolved density of states by cross-section (or matrix-element) effects, which may lead to varying photoemission intensities throughout $k$ space, even if signifcant backfolding did take place.  
Second, the asymmetric spread between antibonding and bonding orbitals around the nonbonding ones \cite{Hoffmann1988} is clearly observed in the experiment. The calculation, however, does not exhibit this effect.  
This is unexpected, as it indicates that the wave functions on the thiophene units in reality could be more extended than predicted by theory. Usually, the opposite is true. 
Nevertheless, the shift of the central energy of the nonbonding quasiband compared to that of the bonding quasiband, introduced by the energy difference of the underlying 1T orbitals, is noticeable in both experiment and calculation. 
Third, only the orbitals contributing to one of the two quasibands exhibit strong intermolecular dispersion. 
The others are almost dispersionless. 
This is strongly correlated with the shape of the real-space wave functions: In the given molecular arrangement, the probability density close to the periphery of the molecules along their long edges clearly favors dispersion.   

Most notably, we found a broad $E(k_x,k_y)$ band with a bandwidth $\sim 0.6$\,eV.
This  band is formed through interactions within and between molecules. 
In one direction ($k_x$), it is based on intermolecular dispersion, in the other ($k_y$) on intramolecular dispersion. 
Because of the finite length of the oligomer, in the intramolecular dispersion direction the band breaks up into six discrete transverse subbands, giving rise to a quasiband in the $k_y$ direction. Interestingly, the inter- and intramolecular dispersions in this band are of similar magnitude, such that  electrons are delocalized nearly isotropically in quasi-one-dimensional strips in each molecular layer.
In addition, we observed experimental hints that the interaction of molecules in the first layer with the substrate leads to further broadening of the $E(k_x,k_y)$ band, in spite of the fact that the metal surface is oxygen-reconstructed, which according to past experience reduces the coupling of flat-adsorbing molecules with the substrate \cite{Yang2018,Wallauer2021,Adamkiewicz2023}. 
Remarkably, the substrate seems to increase both, the intermolecular \textit{and} the intramolecular hopping matrix elements. 

Because of the strong templating effect of the Cu(110)-p($2\times1$)O surface, it is not surprising that the electronic structure of $\alpha$-sexithiophene films is  subject to structure-induced changes when the film thickness is increased. 
It is, however, remarkable that the electronic structure data gathered by POT is sufficient to quantitatively and comprehensively trace the structural changes that occur with growing film thickness. 
This proves the unique potential of POT, not only as a source of high-precision electronic but also structural information. 
That this information is accessible widely in a relatively straightforward in-house angle-resolved ultra-violet photoemission spectroscopy experiment with a momentum microscope further contributes to the attractiveness of POT.

Templating by the Cu(110)-p($2\times1$)O substrate forces $\alpha$-sexithiophene molecules closer to each other than in bulk. 
For steric reasons, they are then also obliged to adopt a steeper angle with respect to the surface plane. 
As the film grows in thickness, both parameters relax towards their bulk values. 
We observe these two effects in two independent experimental quantities: the period of the $E(k_x,k_y)$ band in the intermolecular dispersion direction reveals the intermolecular distance, while the momentum map of a non-dispersing orbital discloses the tilt angle. 
We note that the two measurements, addressing entirely different elements of the electronic structure, independently confirm the thickness-induced changes of the two coupled structure parameters from different vantage points. 
We also observe indications that the steeper tilt angle leads to a stronger intermolecular hybridization, in full accord with expectations based on the structure of the underlying orbitals.

Finally, on the basis of the gathered knowledge about the occupied states, we can venture a prediction regarding unoccupied and excited states in the present system: Because of the nodal structure of its wave function, the LUMO is also expected to exhibit an appreciable intermolecular dispersion, by analogy to the orbitals of the nonbonding quasiband. 
Since the LUMO is not occupied by charge transfer from the substrate (due to the oxygen-induced reconstruction of the surface), it is impossible to test this hypothesis directly. However, when populating the LUMO by photoexcitation of an electron from the HOMO, we can create an exciton in which the hole sits in a localized non-dispersing orbital, while the electron---if our conjecture was correct---would be in a delocalized state. 
Evidently, this delocalization of the electron would have a profound effect on the exciton itself. The system 6T/Cu(110)-p($2\times1$)O in combination with time-resolved POT is therefore an ideal candidate to study exciton delocalization in an organic semiconductor at the level of the exciton wave function~\cite{ExcitonPaper}.\\

\section*{Acknowledgement}
We acknowledge support from the European Union through the Synergy Grant Orbital Cinema (Project ID: 101071259) by the European Research Council (ERC). The computational results have been achieved using the Austrian Scientific Computing (ASC) infrastructure. Further, we thank Dr.\,Anja Haags for fruitful discussions.

\section*{Author contributions}
M.G.R., P.P., and F.S.T. conceived and designed the research. M.S. and E.F. prepared the samples. M.S., A.V.M., E.F., S.S., and F.C.B performed the experiments, and M.S. analyzed the data and made the figures.  S.K. and P.P. did the calculations. F.S.T. supervised the PhD work of M.S. and E.F. The PhD work of S.K. was supervised by P.P. All authors discussed the results. M.S. and F.S.T. wrote the paper with significant input from  S.K., F.C.B, C.K., M.G.R., and P.P.

\section*{Data availability}
The raw experimental data that support the findings of this paper are openly available \cite{ExpData}. The source files for all VASP calculations mentioned in section~\ref{sec:Comp.Details} can be downloaded from the NOMAD repository \cite{nomad6TCuO2025}.

\onecolumngrid
\clearpage
\begin{appendix}
\section{Curvature filter in three dimensions}
\label{Appendix:3D_curvature}

Here, we present a general (dimensionless) expression for the curvature, together with the explicit expression that we used to implement a curvature filter on the $I(E_b,k_x,k_y)$ data cube to generate Fig.\,\ref{fig:3D_Plot}, as discussed in section~\ref{sec:Intermolecular_band}.
For an implicit surface, $F(x,y,z,I) = f(x,y,z) - I = 0$, the curvature is defined as\,\cite{Goldman2005} 
\begin{equation}
\label{Eq:Appendix_curvature_allg}
	K = - \nabla \cdot \left(\frac{\nabla F}{|\nabla F|}\right).
\end{equation}
With $\left(\frac{\partial f}{\partial I}\right) = 0$ and $\nabla \cdot \nabla I = 0$, it follows
\begin{equation}
	K = - \nabla \cdot \left(\frac{\nabla f}{\sqrt{|\nabla f|^2 +1}}\right).
\end{equation}
Calculating the divergence and the gradient yields the following result for the curvature: 

\begin{equation}
	\begin{gathered}
	K = - \Biggl\{\left[1+\left(\frac{\partial f}{\partial {y}}\right)^2+\left(\frac{\partial f}{\partial {z}}\right)^2\right]\frac{\partial^2f}{\partial {x}^{\,2}} + \left[1+\left(\frac{\partial f}{\partial {x}}\right)^2+\left(\frac{\partial f}{\partial {z}}\right)^2\right] \frac{\partial^2f}{\partial {y}^{\,2}} \\ 
	+ \left[1+\left(\frac{\partial f}{\partial {x}}\right)^2+\left(\frac{\partial f}{\partial {y}}\right)^2\right] \frac{\partial^2f}{\partial {z}^{\,2}}	- \frac{\partial f}{\partial {x}}   \frac{\partial f}{\partial {y}}   \left(\frac{\partial^2 f}{\partial {x} \partial {y}} + \frac{\partial^2 f}{\partial {y} \partial {x}}\right)    - \frac{\partial f}{\partial {x}}   \frac{\partial f}{\partial {z}}  \left(\frac{\partial^2 f}{\partial {x} \partial {z}}    + \frac{\partial^2 f}{\partial {z} \partial {x}}\right) \\
	- \frac{\partial f}{\partial {y}}   \frac{\partial f}{\partial {z}} \left(\frac{\partial^2 f}{\partial {y} \partial {z}} + \frac{\partial^2 f}{\partial {z} \partial {y}}\right) \Biggr\}    \Biggl\{1+\left(\frac{\partial f}{\partial {x}}\right)^2 + \left(\frac{\partial f}{\partial {y}}\right)^2 +\left(\frac{\partial f}{\partial {z}}\right)^2 \Biggr\}^{-3/2}.
	\end{gathered}
\end{equation}
The variables $x$, $y$, and $z$ are dimensionless.
However, in the measured data set, there are two independent variables with the same unit ($k_x$ and $k_y$), as well as another independent variable with a different unit ($E$). 
To account for the units, the transformations
\begin{equation}
	\frac{\partial}{\partial x} \rightarrow \xi \frac{\partial}{\partial k_x},~ 
	\frac{\partial}{\partial y} \rightarrow \xi \frac{\partial}{\partial k_y},~ \mathrm{and}~ 
	\frac{\partial}{\partial z} \rightarrow \eta \frac{\partial}{\partial E}
\end{equation}
are used\,\cite{Zhang2011}, where $\xi$ and $\eta$ are constants with the dimensions \AA$^{-1}$ and eV, respectively.
Since the absolute value of the measured intensity, $I = f(E_b,k_x,k_y)$, is only defined to an arbitrary factor, another constant $I_0$ is introduced using the transformation $f \rightarrow I_0 f$\,\cite{Zhang2011}.
The 3D analogue of Eq.~(14) in Ref.\,\cite{Zhang2011} is obtained by defining the parameters $C_k = I_0^2 \xi^2$ and $C_E = I_0^2 \eta^2$.
Note that in Ref.\,\cite{Zhang2011}, $\frac{\partial^2 f}{\partial x \partial y} = \frac{\partial^2 f}{\partial y \partial x} $ is assumed.
This is true as long as the second derivative is continuous (Schwarz theorem)\,\cite{Morino2021}. 
The sign of the curvature depends on the definition of the direction of the surface normal.
For the curvature in three dimensions follows

\begin{equation}
	\label{Eq:Curvature}
	\begin{gathered}
		K \sim - \Biggl\{\left[1+C_k\left(\frac{\partial f}{\partial {k_y}}\right)^2+C_E\left(\frac{\partial f}{\partial {E}}\right)^2\right]C_k\frac{\partial^2f}{\partial {k_x}^{\,2}} + \left[1+C_k\left(\frac{\partial f}{\partial {k_x}}\right)^2+C_E\left(\frac{\partial f}{\partial {E}}\right)^2\right] C_k\frac{\partial^2f}{\partial {k_y}^{\,2}} \\ 
		  + \left[1+C_k\left(\frac{\partial f}{\partial {k_x}}\right)^2+C_k\left(\frac{\partial f}{\partial {k_y}}\right)^2\right] C_E\frac{\partial^2f}{\partial {E}^{\,2}} - C_k^2  \frac{\partial f}{\partial {k_x}}   \frac{\partial f}{\partial {k_y}}  \left( \frac{\partial^2 f}{\partial {k_x} \partial {k_y}} + \frac{\partial^2 f}{\partial {k_y} \partial {k_x}}\right)\\ 
		- C_k\,C_E \frac{\partial f}{\partial {k_x}}   \frac{\partial f}{\partial {E}}  \left(\frac{\partial^2 f}{\partial {k_x} \partial {E}} + \frac{\partial^2 f}{\partial {E} \partial {k_x}}\right) - C_k\,C_E  \frac{\partial f}{\partial {k_y}}   \frac{\partial f}{\partial {E}} \left(\frac{\partial^2 f}{\partial {k_y} \partial {E}} + \frac{\partial^2 f}{\partial {E} \partial {k_y}}\right)\Biggr\} \\
        \times \Biggl\{1+C_k\left(\frac{\partial f}{\partial {k_x}}\right)^2 + C_k\left(\frac{\partial f}{\partial {k_y}}\right)^2 +C_E\left(\frac{\partial f}{\partial {E}}\right)^2 \Biggr\}^{-3/2}. 
	\end{gathered}
\end{equation}

\clearpage
\section{1ML DFT calculation - Additional figure}
\label{Appendix_B}
\begin{figure*}[!htp]
	\centering
	\includegraphics[width=1\textwidth]{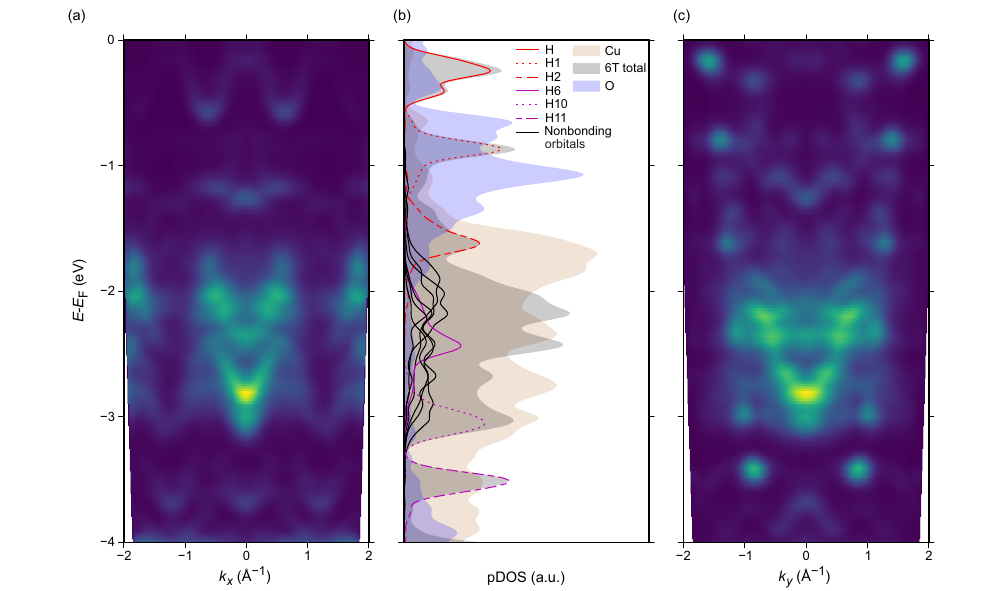}
	\caption{Data of the 1\,ML DFT calculation of Figs.\,\ref{fig:DFT_comparison}\,(a), (d), and (g), displayed on a common energy axis, thus aiding the visual assignment of band to pDOS features.}
	\label{fig:Appendix_1ML_DFT}
\end{figure*}
\end{appendix}
\FloatBarrier

\twocolumngrid
\bibliography{References_6T}

\end{document}